\documentclass[11pt]{article}
\usepackage[affil-it]{authblk}
\usepackage{enumitem} % Enables the use of noitemsep
\usepackage[margin=2cm]{geometry} % Set the page margins
\usepackage{lineno}%\linenumbers
\usepackage{setspace}
\usepackage[utf8]{inputenc}
\usepackage[british]{babel} % or use english instead
\usepackage[scaled]{helvet}
\usepackage[T1]{fontenc}
\usepackage{url}
\usepackage[colorlinks=true, allcolors=blue]{hyperref} % For hyper links
\usepackage{amsmath,amsfonts,amssymb} % For the math functions and fonts
\usepackage{siunitx}
\usepackage{graphicx}
\usepackage{caption}
\usepackage{multicol}
\usepackage{multirow}
\usepackage{threeparttable}
\usepackage[super,sort&compress,numbers]{natbib}
\usepackage{etoolbox}
\AtBeginEnvironment{thebibliography}{\linespread{1}\selectfont}

\begin{document}

\title{Theoretical and numerical modeling of Rayleigh wave scattering\\ by an elastic inclusion}

\author{Shan Li$^{a,b}$, Ming Huang$^b$, Yongfeng Song$^a$, Bo Lan$^{b*}$, Xiongbing Li$^{a}$\footnote{\raggedright{\noindent Corresponding authors. E-mail addresses: lisa\_13@foxmail.com (S. Li), m.huang16@imperial.ac.uk (M. Huang), songyf\_ut@csu.edu.cn (Y. Song), bo.lan@imperial.ac.uk (B. Lan), lixb213@csu.edu.cn (X. Li)}.}}
\affil{
$^a$ School of Traffic and Transportation Engineering, Central South University, Changsha, 410075, China\\
$^b$ Department of Mechanical Engineering, Imperial College London, London SW7 2AZ, United Kingdom
}
\date{\vspace{-5ex}}
\maketitle

\begin{abstract}
This work presents theoretical and numerical models for the backscattering of two-dimensional Rayleigh waves by an elastic inclusion, with the host material being isotropic and the inclusion having arbitrary shape and crystallographic symmetry. The theoretical model is developed based on the reciprocity theorem using the far-field Green's function and the Born approximation, assuming a small acoustic impedance difference between the host and inclusion materials. The numerical finite element (FE) model is established to deliver relatively accurate simulation of the scattering problem and to evaluate the approximations of the theoretical model. Quantitative agreement is observed between the theoretical model and the FE results for arbitrarily-shaped surface/subsurface inclusions with isotropic/anisotropic properties. The agreement is excellent when the wavelength of the Rayleigh wave is larger than, or comparable to, the size of the inclusion, but it deteriorates as the wavelength gets smaller. Also, the agreement decreases with the anisotropy index for inclusions of anisotropic symmetry. The results lay the foundation for using Rayleigh waves for quantitative characterization of surface/subsurface inclusions, while also demonstrating its limitations.
\end{abstract}

\section{\label{sec:1} Introduction}
Rayleigh waves propagating on a solid surface can be scattered by a flaw, such as a void, crack or inclusion, due to the abrupt change of acoustic properties. The scattered waves usually carry information about the geometric and elastic properties of the flaw. Therefore, a thorough understanding of flaw-induced Rayleigh wave scattering is critical for the nondestructive evaluation (NDE) of flaw location and the characterization of flaw size, shape, and other physical characteristics.

Theoretical modeling has been a major method for understanding Rayleigh wave scattering, and most theoretical studies are based on the Kirchhoff and Born approximations. The Kirchhoff approximation is mainly used for scatterers of void or crack type because it treats a scatterer as body sources \cite{ayter1979characterization, achenbach2000calculation,wang2019scattering,phan2013theoretical,phan2013validity,yang2017time, wang2020application}, while the Born approximation uses the incident field to replace the field inside an inclusion \cite{auld1990acoustic,schmerr2016fundamentals} and is therefore suitable for elastic inclusions. The Born approximation has been extensively used for bulk wave scattering \cite{gubernatis1977born, kino1978application} and demonstrated to be effective and accurate when the wavelength is larger, or comparable to, the scatterer size. The Born approximation has also seen applications in studying Rayleigh wave scattering, and example works include those of Auld \cite{auld1979general} and Snieder \cite{snieder19863}. Apart from the Kirchhoff and Born approximations, there have been other theoretical efforts on investigating Rayleigh wave scattering. These include the first-order perturbation theory for weak subsurface inclusions\cite{razin2010scattering}, the reciprocity theorem for linear \cite{schmerr2011ultrasonic} and nonlinear \cite{xu2022surface} Rayleigh waves, and the second-order approximation for multiple shallow cavities \cite{phan2018theoretical}. However, most theoretical studies so far have focused on a specific type of inclusion/flaw, while some  studies involve difficult-to-solve equations \cite{razin2010scattering,schmerr2011ultrasonic}.

The subject of Rayleigh wave scattering has also received considerable numerical studies. Various numerical schemes have been used, and most common ones are the boundary element method and the finite element (FE) method. The boundary element method was mainly used for scatterers that are surface voids (cavities) \cite{arias2004rayleigh, liu2011study, phan2018theoretical}. By comparison, the FE method is capable of dealing with all sorts of scatterers and related examples include its uses in simulating wave propagation \cite{huang2021longitudinal} and scattering \cite{liu2019investigation} in complex polycrystalline media. For this reason, the FE method has been used to analyze the interaction of Rayleigh waves with surface cracks \cite{vu1985diffraction, jian2006surface, hassan2003finite, xu2022surface}, and most recently, it was successfully applied to predicting the attenuation and velocity dispersion of Rayleigh waves in polycrystalline materials \cite{grabec2017numerical,grabec2022surface}. These studies have demonstrated the power of the numerical methods (particularly, the FE method) in realistically simulating Rayleigh wave scattering.

In comparison to the existing studies, this work sets out to study a more general case of Rayleigh wave scattering, with the scatterer being a surface/subsurface elastic inclusion with an arbitrary shape and isotropic/anisotropic property. To achieve this aim, this work contributes to two aspects. First, this work develops a theoretical model based on the reciprocity theorem, utilising the Green's function and the Born approximation, to calculate the backscattering of Rayleigh waves from an arbitrarily-shaped surface/subsurface elastic inclusion with an arbitrary symmetry. Our model is valid for a general elastic inclusion and is not limited to buried scatterers as reported by a similar prior work \cite{snieder19863}. Second, this work also makes use of the proven capability of the FE method to realistically simulate the same scattering problem. This allows for relatively accurate results to be obtained, enabling the validation of our general theoretical model.

The work is organised as follows. Secs. \ref{sec:2} and \ref{sec:3} describe respectively the theoretical and FE models for the backscattering of Rayleigh waves by an elastic inclusion. Sec. \ref{sec:4} compares the results of the theoretical and FE models for a variety of surface and subsurface inclusions of different shapes and elastic anisotropies. Sec. \ref{sec:5} concludes this work.

\section{\label{sec:2} Theoretical model}
\begin{figure}[!ht]
\centering
\includegraphics[width=0.8\textwidth]{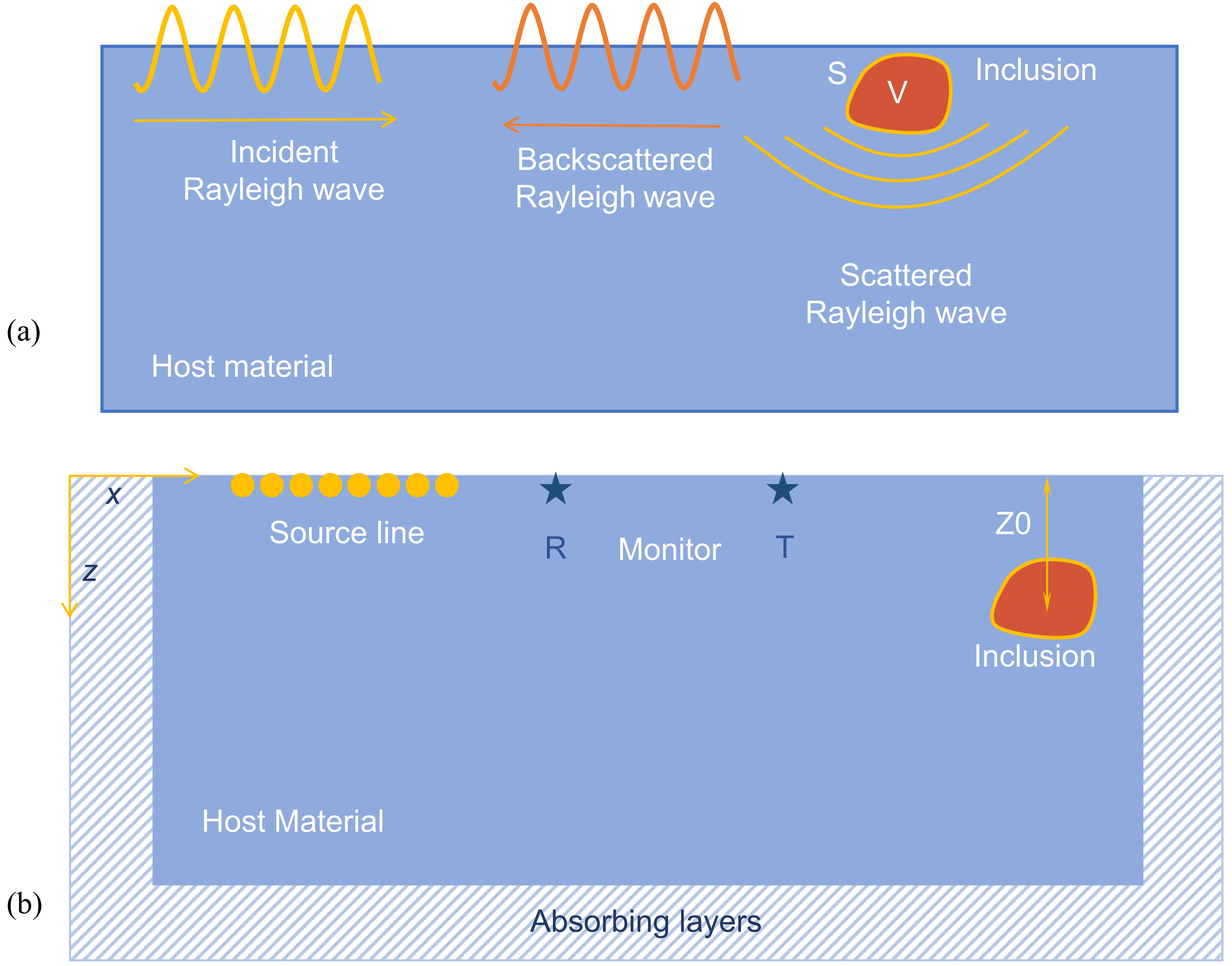}
\caption{\label{fig:1} Schematic of (a) theoretical and (b) finite element models. The yellow dots in (b) are the source line for generating the Rayleigh wave. The originally generated signal is monitored at the transmitting node T and the backscattered signal from the surface/subsurface inclusion A is monitored at the receiving node R. Structured mesh with uniform square elements is used. $Z_0$ is the distance between the upper surface and inclusion. The absorbing boundary is applied on the left, right and bottom of the surface.}
\end{figure}

We consider an isotropic solid with density $\rho_0$ and elastic tensor $c_{pjkl}^0$ in the two-dimensional (2D) half-space defined by the $x-z$ coordinates. As shown in Fig. \ref{fig:1}, an arbitrarily-shaped inclusion is present on the surface or subsurface of the host material. The inclusion is defined in the region $V$ surrounded by the boundary $S$. The inclusion has density $\rho_1\left(\mathbf{x}_{\mathrm{s}}\right)$ and isotropic/anisotropic property described by the elastic tensor $c_{pjkl}^1 \left(\mathbf{x}_{\mathrm{s}}\right)$. A Rayleigh wave propagating in the host material will be scattered as it encounters the inclusion. Two types of scattering arises, one from the incident Rayleigh wave into the same mode and another from the Rayleigh wave into bulk waves. Here we only consider the Rayleigh-to-Rayleigh scattering as it is more prominent than the other type, as proved by prior work \cite{wang2019scattering} and supported by our simulation results in Sec. \ref{sec:3}.

Now we develop a theoretical model to describe the Rayleigh-to-Rayleigh scattering, and we start from the reciprocity theorem that gives the scattered Rayleigh wave by\cite{schmerr2011ultrasonic,schmerr2016fundamentals}
\begin{equation}
\begin{split}
\label{eq:001}
{u}^\text{sc}_{n} \left(\mathbf{x}, \omega \right) & = \int_S [c_{pjkl}^0 n_l \left( \mathbf{x}_{\mathrm{s}} \right) {u}_k^+\left( \mathbf{x}_{\mathrm{s}},\omega \right)  G_{nj,p}\left(\mathbf{x}, \mathbf{x}_{\mathrm{s}},\omega \right)  \\ & \quad \quad- c_{pjkl}^0 n_p \left( \mathbf{x}_{\mathrm{s}}\right)  G_{nj}\left(\mathbf{x}, \mathbf{x}_{\mathrm{s}},\omega \right)  {u}_{k,l}^+\left( \mathbf{x}_{\mathrm{s}},\omega \right) ]   \mathrm{~d}  S  \textrm{,}
\end{split}
\end{equation}
where the Einstein summation convention over repeated indices ($p$, $j$, $k$, $l$) from 1 to 3 (or $x$ to $z$) is assumed. $\mathbf{x}$ is the point of interest at which we evaluate the scattered wave, while $\mathbf{x}_{\mathrm{s}}$ is a point in the inclusion $V$. ${u}_{l}(\mathbf{x}_{\mathrm{s}})$ is the $l$-th displacement component of the Rayleigh wave. $G_{nj}(\mathbf{x}, \mathbf{x}_{\mathrm{s}})$ is the dyadic representation of the Green's function. $n_{k}(\mathbf{x}_{\mathrm{s}})$ is the $k$-th component of the outward unit normal to the inclusion surface. The comma derivative notation is used and the derivative is over the scattered coordinate throughout this work; e.g. $ {u}_{k,l}= \partial {u}_k/\partial x_{\mathrm{s}l}$ with $x_{\mathrm{s}l}$ being the $l$-th component of $\mathbf{x}_{\mathrm{s}}$. The plus and minus superscripts are used to indicate quantities that are evaluated on the host or inclusion side of the surface $S$, respectively. From the continuity of displacement and traction across $S$, we have
\begin{equation}
\label{eq:002}
   {u}_k^+\left( \mathbf{x}_{\mathrm{s}},\omega \right) = {u}_k^-\left( \mathbf{x}_{\mathrm{s}},\omega \right) \textrm{,} \quad
   n_p c_{pjkl}^0  {u}_{k, l} ^{+} \left( \mathbf{x}_{\mathrm{s}},\omega \right)  = n_p c_{pjkl}^1  {u}_{k, l}^-\left( \mathbf{x}_{\mathrm{s}},\omega \right) \textrm{,}
\end{equation}
so Eq. \ref{eq:001} becomes
  \begin{equation}
   \begin{split}
      {u}^\text{sc}_{n} \left(\mathbf{x}, \omega \right)  & = \int_S [c_{pjkl}^0 n_l \left( \mathbf{x}_{\mathrm{s}}\right) {u}_k ^ {-} \left( \mathbf{x}_{\mathrm{s}},\omega \right) \partial G_{nj, p}\left( \mathbf{x}, \mathbf{x}_{\mathrm{s}},\omega \right)   \\ & \quad \quad - c_{pjkl}^1\left(\mathbf{x}_{\mathrm{s}}\right) n_p \left( \mathbf{x}_{\mathrm{s}}\right)  G_{nj}\left(\mathbf{x}, \mathbf{x}_{\mathrm{s}},\omega \right)  {u}_{k,l}^{-} \left( \mathbf{x}_{\mathrm{s}},\omega \right)  ]   \mathrm{~d} S \textrm{.}
\label{eq:003}
  \end{split}
 \end{equation}
Applying the divergence theorem $\int_{V}  v_{i,k} d V = \int_{S} v_{i} n_{k} d S $ to the right side of Eq. \ref{eq:003} leads to

\begin{equation}
\begin{split}
 {u}^\text{sc}_{n} \left(\mathbf{x}, \omega \right)  & = \int_{V} \{  \left[ c_{pjkl}^0  {u}_k \left( \mathbf{x}_{\mathrm{s}},\omega \right)  G_{nj, p}\left(\mathbf{x}, \mathbf{x}_{\mathrm{s}},\omega \right)  \right]\textrm{,}_{l} \\
 & \quad \quad -   \left[c_{pjkl}^1 G_{nj}\left(\mathbf{x}, \mathbf{x}_{\mathrm{s}},\omega \right)  {u}_{k, l} \left( \mathbf{x}_{\mathrm{s}},\omega \right) \right]\textrm{,}_{p} \}  \mathrm{~d} V \textrm{,}
\label{eq:004}
\end{split}
\end{equation}
where we have dropped the minus superscript since the integration points are all within the inclusion $V$. Now substituting the equation of motion\cite{schmerr2016fundamentals}
\begin{equation}
\begin{split}
\left[ c_{pjkl}^0   G_{nj,p}\left(\mathbf{x}, \mathbf{x}_{\mathrm{s}},\omega \right)  \right]\textrm{,}_{k} +\rho_0\omega^2 G_{nl} \left(\mathbf{x}, \mathbf{x}_{\mathrm{s}},  \omega\right) &= -\delta_{nl}\delta \left(\mathbf{x}_{\mathrm{s}}-\mathbf{x} \right) \textrm{,} \\
\left[ c_{pjkl}^1 \left( \mathbf{x}_{\mathrm{s}},\omega \right)  {u}_{k,l} \left( \mathbf{x}_{\mathrm{s}},\omega \right)  \right]\textrm{,}_{p} &=-\rho_1 \omega^2 {u}_j \left(\mathbf{x}_{\mathrm{s}}, \omega \right) \textrm{,}
\label{eq:005}
\end{split}
\end{equation}
into Eq. \ref{eq:004} and using the sampling properties of the delta function, we have 
\begin{equation}
\begin{split}
   {u}^\text{sc}_{n} \left(\mathbf{x}, \omega \right)  = \int_{V} &\left [ \omega^2 \Delta \rho G_{nl} \left(\mathbf{x},\mathbf{x}_{\mathrm{s}}, \omega \right) {u}_{l} \left(\mathbf{x}_{\mathrm{s}}, \omega \right) \right.\\
   &\left. -\Delta c_{pjkl}(\mathbf{x}_{\mathrm{s}})  G_{nj, p} \left(\mathbf{x},\mathbf{x}_{\mathrm{s}}, \omega \right) {u}_{k, l} \left(\mathbf{x}_{\mathrm{s}}, \omega \right) \right ]  \mathrm{~d} V \textrm{,}
\label{eq:006}
\end{split}
\end{equation}
with
\begin{equation}
\begin{gathered}
\begin{alignedat}{2}
\Delta \rho &= \rho_1(\mathbf{x}_{\mathrm{s}})-\rho_0 \quad \textrm{and} \quad \Delta c_{pjkl}(\mathbf{x}_{\mathrm{s}}) &= c_{pjkl}^1(\mathbf{x}_{\mathrm{s}}) - c_{pjkl}^0  \textrm{.}
\end{alignedat}
\end{gathered}
\end{equation}

Equation \ref{eq:006} provides an exact solution for the scattering amplitude, but the solution is intractable. We address this difficulty by invoking the Born approximation with assuming a small property difference between the inclusion and host materials. Here we do not attempt to define how small the property difference needs to be for obtaining a reasonably accurate solution, but as we shall see in Sec. \ref{sec:4}, it depends on various factors such as frequency, inclusion size and material anisotropy. As a result of the Born approximation, the incident wave is only slightly perturbed by the inclusion, and therefore the displacement and its derivative in Eq. \ref{eq:006} can be obtained from their values due to the incident wave only \cite{schmerr2016fundamentals}, i.e.
\begin{equation}
\begin{gathered}
{u}_{l}(\mathbf{x}_{\mathrm{s}}) \approx {u}_{l}^\text{in}(\mathbf{x}_{\mathrm{s}}) \quad \textrm{and} \quad  {u}_{k, l}(\mathbf{x}_{\mathrm{s}}) \approx {u}_{k, l}^\text{in}(\mathbf{x}_{\mathrm{s}}) \textrm{.}
\end{gathered}
\end{equation}

The formulation so far is valid for both two-dimensional or three-dimensional cases. Since the 3D dyadic Green’s function is not known yet, here we only address the 2D case. Now, we assume that the incident Rayleigh wave is a time-harmonic plane wave propagating along the surface in the $x$-direction, as shown in Fig. \ref{fig:1}(a). In this case, the unit displacement components and their derivatives can be given by \cite{auld1990acoustic,rose2000ultrasonic, phan2013application}
\begin{equation}
\label{eq:8}
\begin{split}
 {u}_{k}^\text{in}(\mathbf{x}_{\mathrm{s}}) = {d}_{k}^\text{in}(z_\mathrm{s})\exp{\left(\text{i} k_R \mathbf{e}^\text{in} \cdot \mathbf{x}_{\mathrm{s}} \right)}\textrm{,} &  \quad   {u}_{k, l}^\text{in}(\mathbf{x}_{\mathrm{s}}) = \left[{d}_{k, l}^\text{in}(z_\mathrm{s}) + \text{i} {d}_{k}^\text{in}(z_\mathrm{s}) k_R {e}^\text{in}_{l} \right] \exp{\left(\text{i} k_R \mathbf{e}^\text{in} \cdot \mathbf{x}_{\mathrm{s}} \right)}\textrm{,}\\
\end{split}
\end{equation}
with
\begin{equation}
\begin{split}
  &{d}^\text{in}_{k}  = \left[U_{R}(z_\mathrm{s}) , 0 , \text{i}  W_{R}(z_\mathrm{s})\right]\textrm{,}  \\
   U_{R}(z_\mathrm{s})    & =  \frac{k_R}{p}  \frac{(2 c_T^2-c_R^2)}{2 c_T^2 } \exp(-p z_\mathrm{s}) - \frac{q}{k_R}   \exp(-q z_\mathrm{s}) \textrm{,}\\
    W_{R}(z_\mathrm{s})   & = \frac{(2 c_T^2-c_R^2)}{2 c_T^2 }  \exp(-p z_\mathrm{s})- \exp(-q z_\mathrm{s}) \textrm{,}
    \end{split}
\end{equation}
where $p = k_R \sqrt{1-c_R^2/c_L^2} $  and $q = k_R \sqrt{1-c_R^2/c_T^2} $. $\mathbf{e}^\text{in} = [1,0,0]$ is the propagation direction of the incident Rayleigh wave. $k_R$ and $c_R$ are the wave number and phase velocity of the incident Rayleigh wave. The phase velocity $c_R$ can be calculated by
\begin{equation}
\label{Eq:cr}
 (2-c_R^2/c_T^2)^2-4(1-c_R^2/c_L^2)^{1/2}(1-c_R^2/c_T^2)^{1/2} = 0 \textrm{,}
\end{equation}
where $c_L$ and $c_T$ are the velocities of the longitudinal and shear waves in the host material.

For the 2D case considered here, the dyadic Green's function is given by \cite{snieder19863,aki2002quantitative}
\begin{equation}
    \begin{split}
        \label{eq:010}
        G_{nl} \left(\mathbf{x}, \mathbf{x}_{\mathrm{s}}, \omega \right)  =A_0 \frac{\exp(\text{i} k_R r_2)}{\sqrt[]{2\pi r_2/k_R}} [{d}_{l}^\text{sc} \left(z_\mathrm{s} \right)]^{*} {p}^\text{sc}_{n} \left(z \right) \textrm{,}
    \end{split}
\end{equation}
with
\begin{equation}
    \begin{split}
         & \quad \quad \quad  A_0  =  \frac{1}{4P_{R}c_R}\frac{\exp(\text{i} \pi/4)}{ k_R}, \quad \\
         {p}^\text{sc}_{n} \left(z \right) = & \left[ U_{R}\left(z \right) ,0, \text{i} W_{R} \left(z \right)  \right] \textrm{,} \quad {d}_{l}^\text{sc}\left(z_\mathrm{s} \right) = \left[-U_{R}(z_\mathrm{s}) , 0 , \text{i}  W_{R}(z_\mathrm{s})\right]\textrm{,} \\ &P_R = \frac{1}{2} \rho_0 {c}_{g} \int_o ^{\infty } \left[ U_{R}\left(z\right)^2+ W_{R}\left(z\right)^2 \right]\mathrm{~d} z       \textrm{,}
         \end{split}
\end{equation}
where $P_R$ represents a normalized power per unit width in the travelling wave mode. ${c}_{g}$ is the group velocity of the Rayleigh wave. The asterisk superscript denotes the complex conjugate. In the far field, the distance $r_2$  between the evaluation point $\mathbf{x}$ and a point $\mathbf{x}_{\mathrm{s}}$ on the inclusion can be approximated as \cite{schmerr2011ultrasonic,schmerr2016fundamentals}
\begin{equation}
r_2 =  \mathbf{x}- \mathbf{x_\mathrm{s}} \approx r-\mathbf{e}^\text{sc} \cdot \mathbf{x}_{\mathrm{s}}  
\end{equation}
where  a fixed point on the inclusion is taken as the origin of the $\left(x, z \right)$ coordinates. $\mathbf{e}^\text{sc}$ is the 2D unit vector in the $\left(x,z \right)$ plane from the origin $O$ to point $\mathbf{x}$, representing the propagation direction of the scattered Rayleigh wave. $r$ is the distance between $O$ and $\mathbf{x}$, i.e., $r = \sqrt{x^2+z^2}$. 

The Green's function in Eq. \ref{eq:010} applies to any scattering direction. For simplicity, we only consider the scattering in the backward direction of the incident wave, namely $\mathbf{e}^\text{sc}=[-1,0,0]$. Then, the Green's function and its derivative can be written as
\begin{equation}\label{eq:green}
\begin{split}
        G_{nl} \left(\mathbf{x}, \mathbf{x}_{\mathrm{s}}, \omega \right)   & = A_0 \frac{\exp(\text{i} k_R r)}{\sqrt[]{r}} [{d}_l^\text{sc} \left(z_\mathrm{s} \right)]^{*} {p}^\text{sc}_{n} \left(z \right) \exp(-\text{i} k_R \mathbf{e}^\text{sc}\cdot \mathbf{x}_{\mathrm{s}})\textrm{,} \\
        G_{nj,p} \left(\mathbf{x},\mathbf{x}_{\mathrm{s}}, \omega \right)   = & A_0 \frac{\exp(\text{i} k_R r)}{\sqrt[]{r}} \left\{ [{d}_{j,p}^\text{sc} \left(z_\mathrm{s} \right)]^{*} -\text{i} k_R {e}^\text{sc}_p [{d}_j^\text{sc} \left(z_\mathrm{s} \right)]^{*} \right\}   \\
        & \quad \times {p}^\text{sc}_{n} \left(z \right)  \exp{\left(-\text{i} k_R \mathbf{e}^\text{sc}\cdot \mathbf{x}_{\mathrm{s}}\right)}\textrm{.}
\end{split}
\end{equation}

Substituting Eq. \ref{eq:green} into Eq. \ref{eq:006} and rearranging the result, we have 
\begin{equation}
{u}^\text{sc}_{n}\left(\mathbf{x}, \omega \right)  = A^\text{sc} \left(\omega \right) \frac{\exp(\text{i} k_R r)}{\sqrt[]{2\pi r/k_R}} {p}^\text{sc}_{n} \left(z \right) \textrm{,}
\end{equation}
where $A^\text{sc} \left( \omega \right)$ is the far-field amplitude of the backscattered Rayleigh wave, given by
\begin{equation}
\label{eq:016}
\begin{split}
        A^\text{sc}\left( \omega \right)  = A_0  \int_{V}  & \left\{\Delta \rho \omega^2  d^\text{in}_{l} \left(\mathbf{x}_{\mathrm{s}}\right) [{d}^\text{sc}_{l} \left(\mathbf{x}_{\mathrm{s}} \right)]^{*} - \Delta c_{pjkl}(\mathbf{x}_{\mathrm{s}}) N(\mathbf{x}_{\mathrm{s}}) M(\mathbf{x}_{\mathrm{s}}) \right\}  \\ & \times \exp{\left[\text{i} k_R \left(\mathbf{e}^\text{in}- \mathbf{e}^\text{sc}\right)\cdot \mathbf{x}_{\mathrm{s}} \right]} \mathrm{~d} V \textrm{,}
\end{split}
\end{equation}
with
\begin{equation}
  N(\mathbf{x}_{\mathrm{s}}) ={d}_{k, l}^\text{in}\left(z_\mathrm{s} \right) + \text{i}k_R {e}^\text{in}_{l} {d}_{k}^\text{in}\left(z_\mathrm{s} \right) \quad \textrm{and} \quad M(\mathbf{x}_{\mathrm{s}}) = [{d}_{j,p}^\text{sc} \left(z_\mathrm{s} \right)]^{*} -\text{i} k_R {e}^\text{sc}_p [{d}_j^\text{sc} \left(z_\mathrm{s} \right)]^{*} \textrm{.}
\end{equation}
Eq. \ref{eq:016} can be further evaluated and the resulting final expression is provided in Appendix A. Note that Einstein summation convention over repeated indices ($p$, $j$, $k$, $l$) is only for 1 and 3 (or $x$ to $z$) in the 2D case.

Although Eq. \ref{eq:016} is derived for isotropic inclusions, it can be modified to accommodate anisotropic inclusions if the root-mean-square (RMS) of the backscattering response from a lot of randomly-oriented inclusions is concerned. Let us consider a case that the host material is the Voigt average of the anisotropic inclusion, with the elastic constants given by
\begin{equation}\label{eq:voigt}
\begin{split}
 c_{11}^0=\left<c_{11}\right> &= \frac{3(c_{11}+c_{22}+c_{33})+2(c_{23}+c_{13}+c_{12})+4(c_{44}+c_{55}+c_{66})}{15} \\
 c_{44}^0=\left<c_{44}\right> &= \frac{(c_{11}+c_{22}+c_{33})-(c_{23}+c_{13}+c_{12})+3(c_{44}+c_{55}+c_{66})}{15} \textrm{,}
\end{split}
\end{equation}
 The fourth-rank elastic tensor $c_{pjkl}$ is written as $c_{ij}$ using the Voigt index notation where the pairs of indices are contracted to the following single values: $11\rightarrow 1$, $22\rightarrow 2$, $33\rightarrow 3$, $23\,\mathrm{or}\,32\rightarrow 4$, $13\,\mathrm{or}\,31\rightarrow 5$ and $12\,\mathrm{or}\,21\rightarrow 6$. The host material has the same density as the inclusion, namely $\rho_0=\rho_1$. In this case, the RMS of the backscattering amplitude from an infinite number of inclusions with random crystallographic orientations can be calculated by
\begin{equation}\label{eq:020}
    \begin{split}
       A^\text{sc}_{rms}\left(\omega \right) = \sqrt{\left< A_{pjkl}^\text{sc}\left(\omega \right) A^{sc}_{\alpha\beta\gamma\delta}\left(\omega \right)\right>} = \sqrt{\left< \sum _{pjkl}^{1, 3} A_{pjkl}^\text{sc}\left(\omega \right) \sum_{\alpha\beta\gamma\delta}^{1, 3} A^{sc}_{\alpha\beta\gamma\delta}\left(\omega \right) \right>} \textrm{.}
    \end{split}
\end{equation}

The developed theoretical model calculates the far-field amplitude of the backscattered Rayleigh wave from an arbitrarily-shaped inclusion in 2D. The size, shape and depth of the inclusion, and its property contrast to the host material are all incorporated in the integral in Eqs. \ref{eq:016} and \ref{eq:020}. For the inclusions considered in this work, the integral is evaluated by numerical integration for both regularly and irregularly shaped inclusions. A major advantage of the model is that it is capable of dealing with arbitrarily-shaped surface/subsurface inclusions of isotropic or general anisotropic properties. The use of the Born approximation, however, means that the property contrast between the inclusion and the host material needs to be necessarily small; thus, the traditional Kirchhoff approximation is preferable when the scatterer is a void or crack. We will use numerical simulations to evaluate how the Born approximation affects the accuracy of the obtained solution.

\section{\label{sec:3}Finite element model}

The FE method has been demonstrated recently to be powerful and accurate for simulating the propagation and scattering of bulk \cite{van2015finite,van2017finite,van2018numerical,Ryzy2018,Bai2018,liu2019investigation,huang2020maximizing} and surface \cite{sarris2021attenuation,grabec2022surface} waves in complex solids. Here we report our use of the method for simulating the same physical problem as addressed by the above theoretical model, which is the backscattering of a Rayleigh wave by a surface/subsurface inclusion. Note that we are concerned with the wave phenomena in the middle of a wide wavefront of a propagating Rayleigh wave, and the symmetry means that this can be simplified and well simulated by a 2D FE model. 

As schematically shown in Fig. \ref{fig:1}, the 2D FE model is based in the $x-z$ plane. The dimension of the model depends on the modelling frequency and inclusion type, and detailed parameters are given in Table \ref{tab:model} for the modeled cases. The model space is discretized with uniform linear square elements, with an edge size $h$ of about one sixtieth of the center-frequency wavelength of the Rayleigh wave in the host material to minimize numerical error to $\sim0.1\%$ \cite{van2015finite,van2017finite,huang2020maximizing}. The inclusion is modelled by an aggregate of elements that are assigned with the density $\rho_1$ and elastic tensor $c_{pjkl}^1$ of the inclusion material. The remaining elements in the middle of the model are defined as the host material with density $\rho_0$ and elastic tensor $c_{pjkl}^0$. The inclusions considered in this work are summarized in Table \ref{tab:inclusion}. For isotropic inclusions, the host material is defined as aluminum with Young's modulus $E_0$ = $70$ GPa, Poisson's ratio $\nu_0$ = $0.35$ and density $\rho_0$ = 2700 kg/m$^3$, and the inclusion has 4\% impedance contrast to the host material caused by density or/and Young's modulus differences. For anisotropic inclusions, the host material has the same density as the inclusion (i.e., $\rho_0=\rho_1$), and its isotropic elastic tensor $c_{pjkl}^0$ is the Voigt average of $c_{pjkl}^1$ calculated by Eq. \ref{eq:voigt}. The anisotropic inclusion materials considered in this work are listed in Table \ref{tab:3}.

\begin{table}[t]
\renewcommand{\arraystretch}{1.5}
\setlength\tabcolsep{10pt}
\small
\centering
\caption{\label{tab:model} Models used in the simulation. Center frequency of FE modelling $f_c$ (MHz), dimensions $d_x \times d_z$  (mm $\times$ mm), mesh size $h$ (mm), degree of freedom (d.o.f); copper sulfate pentahydrate is abbreviated as CSP.}
\begin{tabular}{cccccc}
\hline\hline
\multirow{1}{*}{Material} & \multirow{1}{*}{ $f_c$} & \multirow{1}{*}{$d_x \times d_z$}  & \multirow{1}{*}{$h$}  & \multirow{1}{*}{d.o.f}
\\ \hline
\multirow{2}{*}{Aluminum}                                   &  $0.5$     & 560  $\times $ 56                        &  96 $\times$ 10$^{-3}$                 &       6.8 $\times$ 10$^{6}$   \\ 
 &  $1$     & 280  $\times $ 28                         &  48 $\times$ 10$^{-3}$                 &       6.8 $\times$ 10$^{6}$                                             \\\hline 
Aluminum               & $2$       & 140  $\times $ 14                        &   24 $\times$ 10$^{-3}$            &    6.8 $\times$ 10$^6$    
\\ 
Inconel, Lithium                                                  & $2$       & 160  $\times $ 16                        &   24 $\times$ 10$^{-3}$            &    8.9 $\times$ 10$^6$   & 
                     \\ 
CSP                                            & $2$       & 120  $\times $ 12                        &   18 $\times$ 10$^{-3}$            &    8.9 $\times$ 10$^6$    
\\ 
\hline
Aluminum                                             & $4$       & 70  $\times $ 7                        &   12 $\times$ 10$^{-3}$            &    6.8 $\times$ 10$^6$                                                   \\
Inconel, Lithium                                            & $4$       & 80  $\times $ 8                        &   12 $\times$ 10$^{-3}$            &    8.9 $\times$ 10$^6$                                                   \\
 CSP                                    & $4$       & 60  $\times $ 6                        &   9 $\times$ 10$^{-3}$            &    8.9 $\times$ 10$^6$                                                 \\ \hline
Aluminum                                          & $8$       & 35  $\times $ 3.5                        &   6 $\times$ 10$^{-3}$            &    6.8 $\times$ 10$^6$                                                   \\
Inconel, Lithium                                       & $8$       & 40  $\times $ 4                        &   6 $\times$ 10$^{-3}$            &    8.9 $\times$ 10$^6$     \\ 
CSP                                                    & $8$       & 30  $\times $ 3                        &   4.5 $\times$ 10$^{-3}$            &    8.9 $\times$ 10$^6$        \\
\hline 
\multirow{3}{*}{Aluminum}                                                  & $16$       & 17.5  $\times $ 1.75                        &   3 $\times$ 10$^{-3}$            &    6.8 $\times$ 10$^6$                                            \\ 
 & $20$       & 14  $\times $ 1.4                        &   2.5 $\times$ 10$^{-3}$            &    6.8 $\times$ 10$^6$    &                                               \\ 
                & $32$       & 8.75 $\times $ 0.875                        &   1.5 $\times$ 10$^{-3}$            &    6.8 $\times$ 10$^6$                                                 \\ 
\hline\hline
\end{tabular}
\end{table}

\begin{table}[t]
\renewcommand{\arraystretch}{1.5}
\setlength\tabcolsep{10pt}
\small
\centering
\caption{\label{tab:inclusion} Inclusion shape and size used in the simulations. A regularly-shaped inclusion has depth $D$ (mm) and width $L$ (mm), and an irregular inclusion has an equivalent radius, $d=\sqrt{s/\pi}$, with $s$ being the area of the inclusion.}
\begin{tabular}{llccccccc}
\hline\hline
 &
\multicolumn{1}{c}{Inclusion} & \multicolumn{1}{c}{Illustration} &
\multirow{1}{*}{$D$}  & \multirow{1}{*}{$L$}  &  \multirow{1}{*}{ $d$}  &\multirow{1}{*}{ $Z_0/ \lambda$}  &\\ \hline
\multirow{5}{*}{Regular} & Half-circle 
&\parbox[c]{0.1\textwidth}{\includegraphics[width=0.08\textwidth]{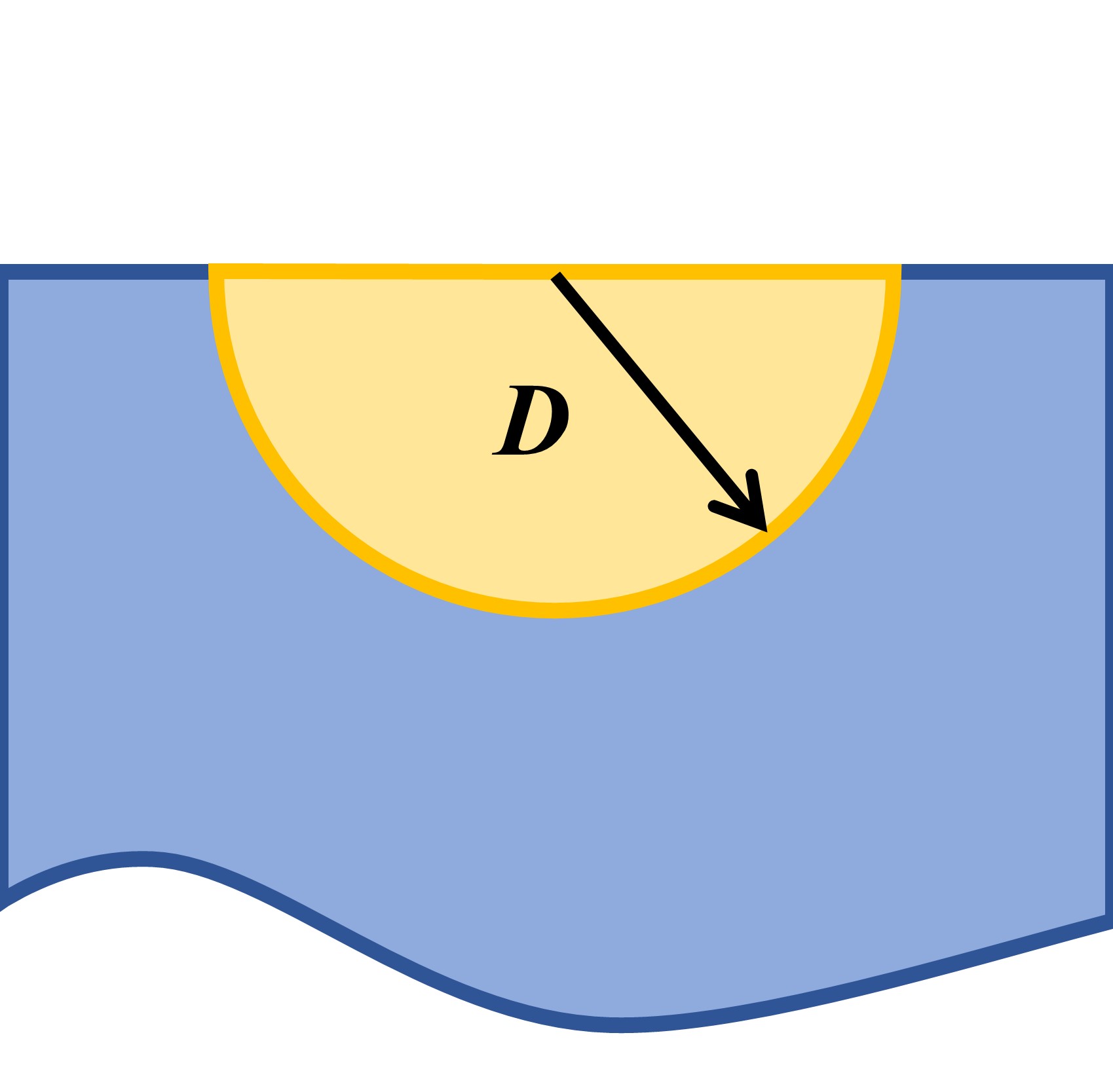}}  
            & \multirow{4}{*}{$0.19$}                        &  $D$                       &    -   &\multirow{4}{*}{0}    \\ 
 & half-ellipse             &\parbox[c]{0.1\textwidth}{\includegraphics[width=0.08\textwidth]{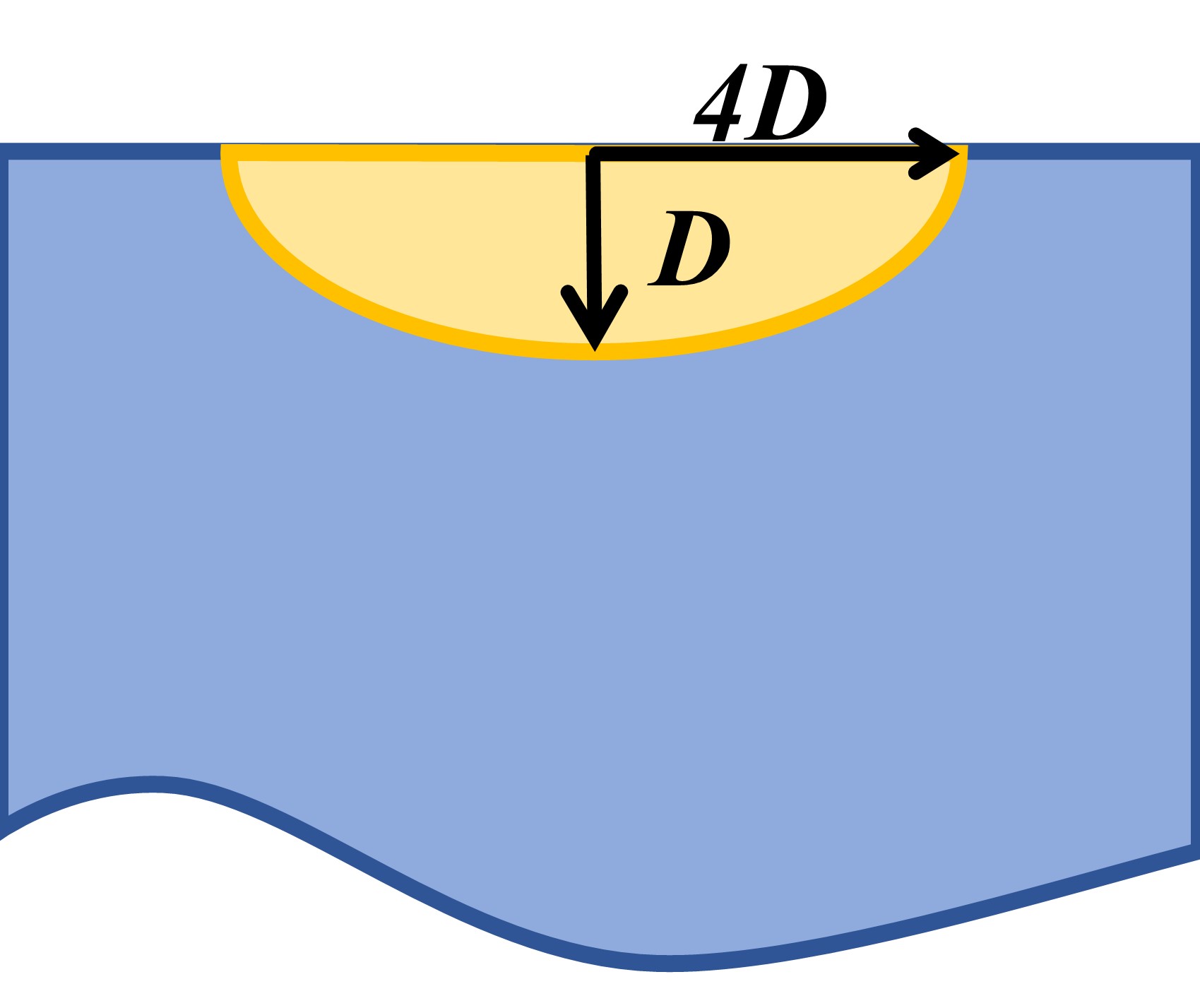}}                                             &                         &   4$D$     &    -                      \\
 & Square            &\parbox[c]{0.1\textwidth}{\includegraphics[width=0.08\textwidth]{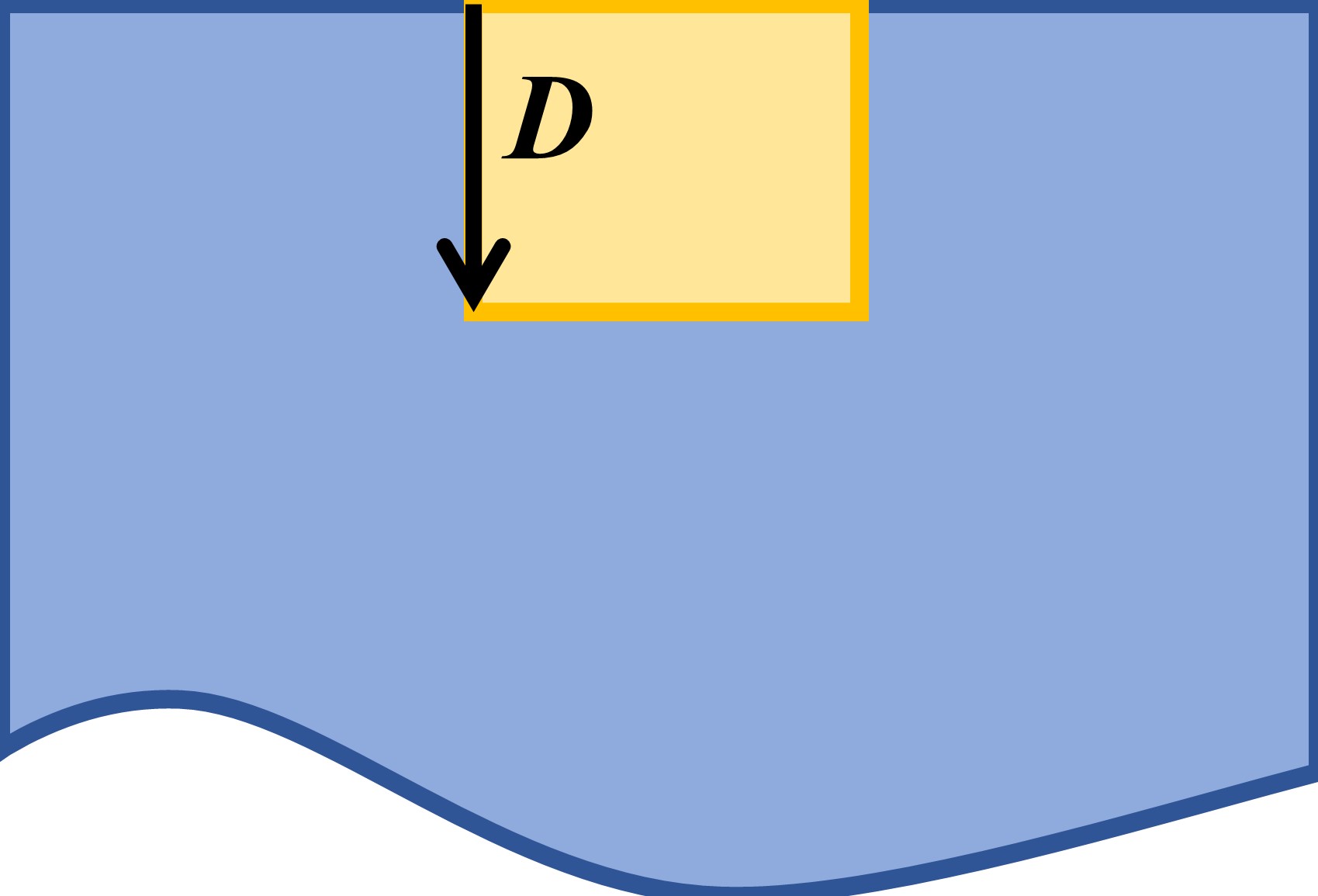}}                                           &                       &  $D$             &   -   \\ 
 & Rectangle                &\parbox[c]{0.1\textwidth}{\includegraphics[width=0.08\textwidth]{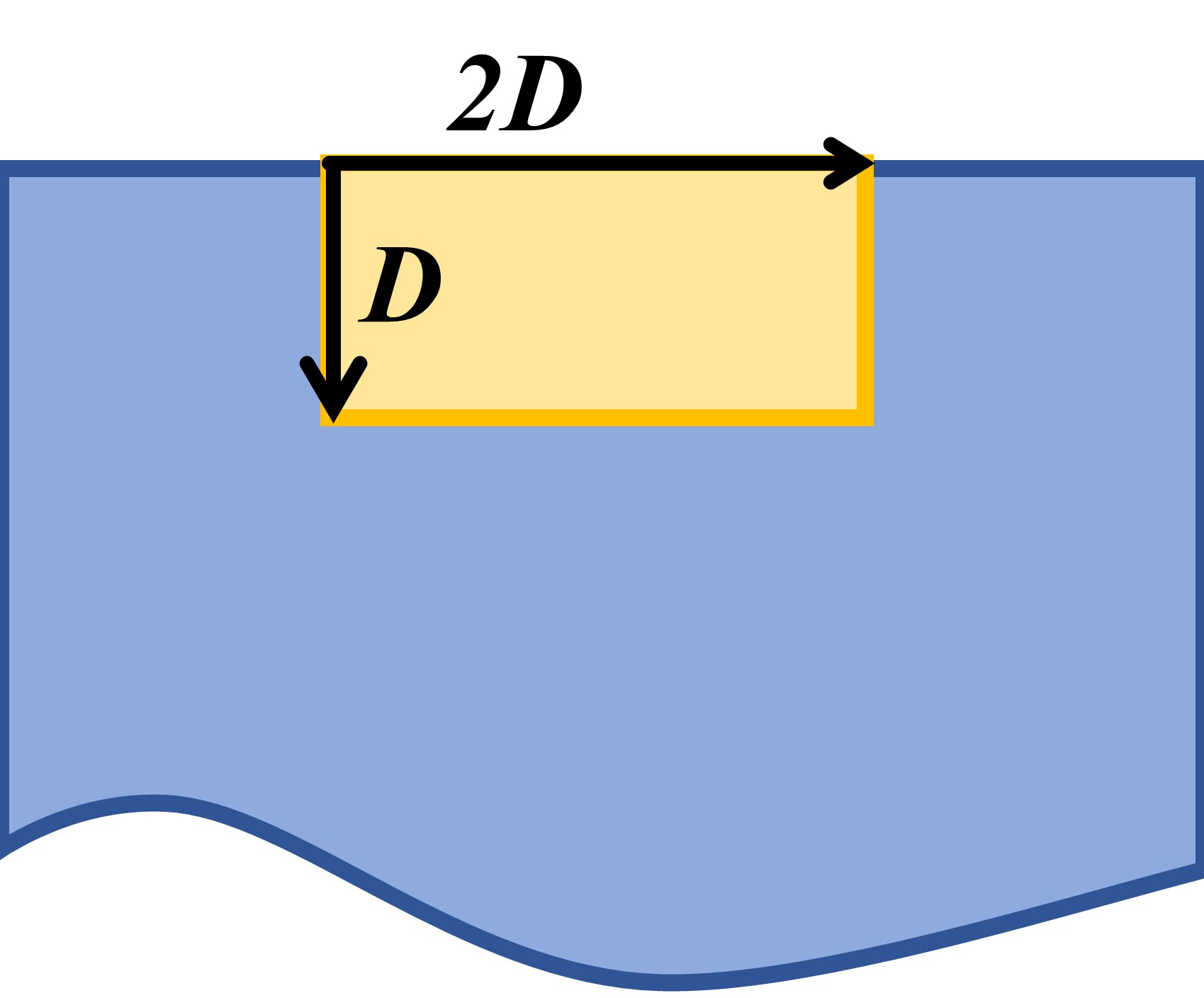}}                                         &                       &  2$D$    &- 
 \\ \cline{2-8}
 & Circle       
&\parbox[c]{0.1\textwidth}{\includegraphics[width=0.08\textwidth]{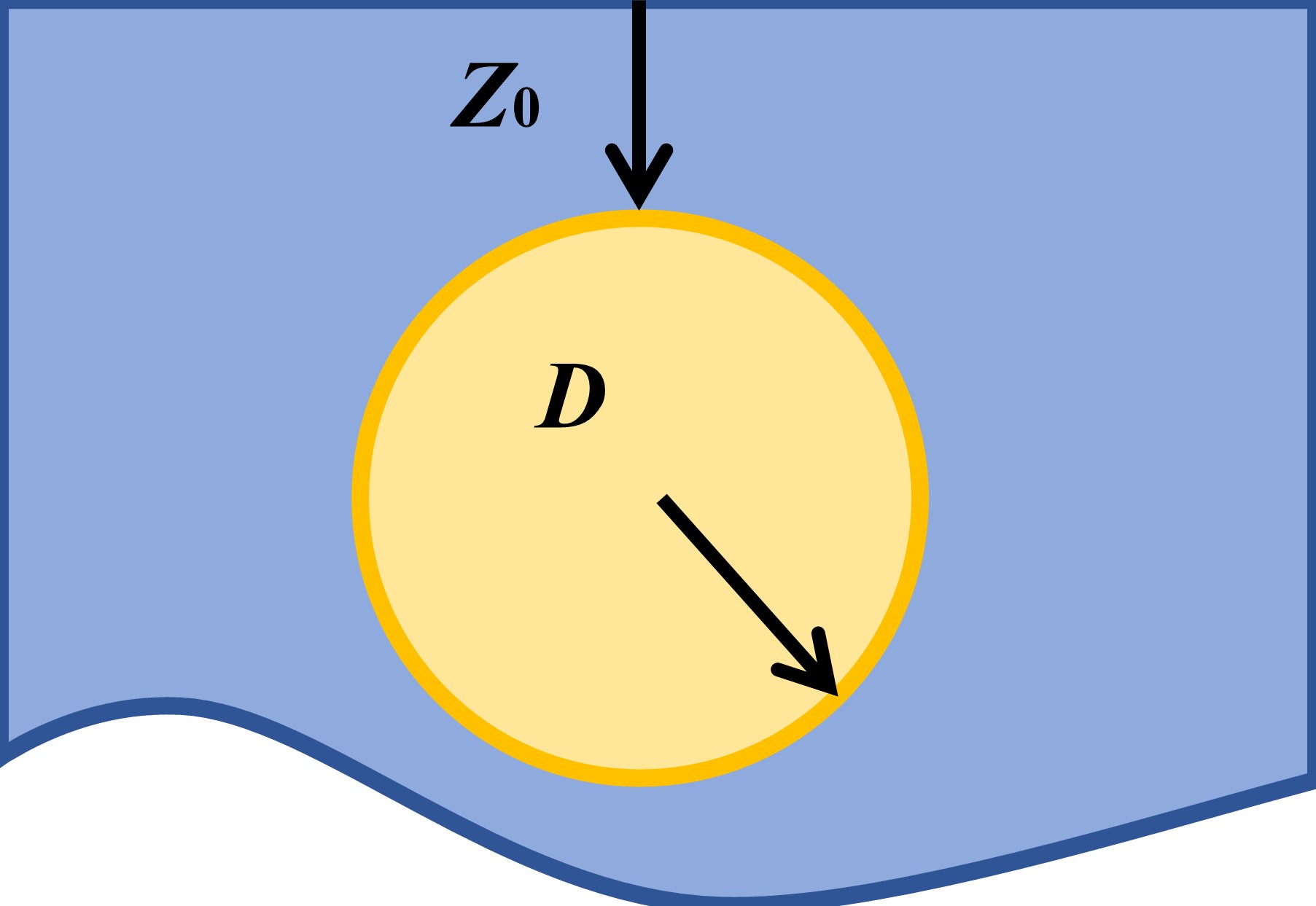}}
 & 0.38                       &  $D$     &- & 0 $\sim$ 2
\\ \hline
\multirow{1}{*}{Irregular}
 & Inclusion $M$            &\parbox[c]{0.1\textwidth}{\includegraphics[width=0.08\textwidth]{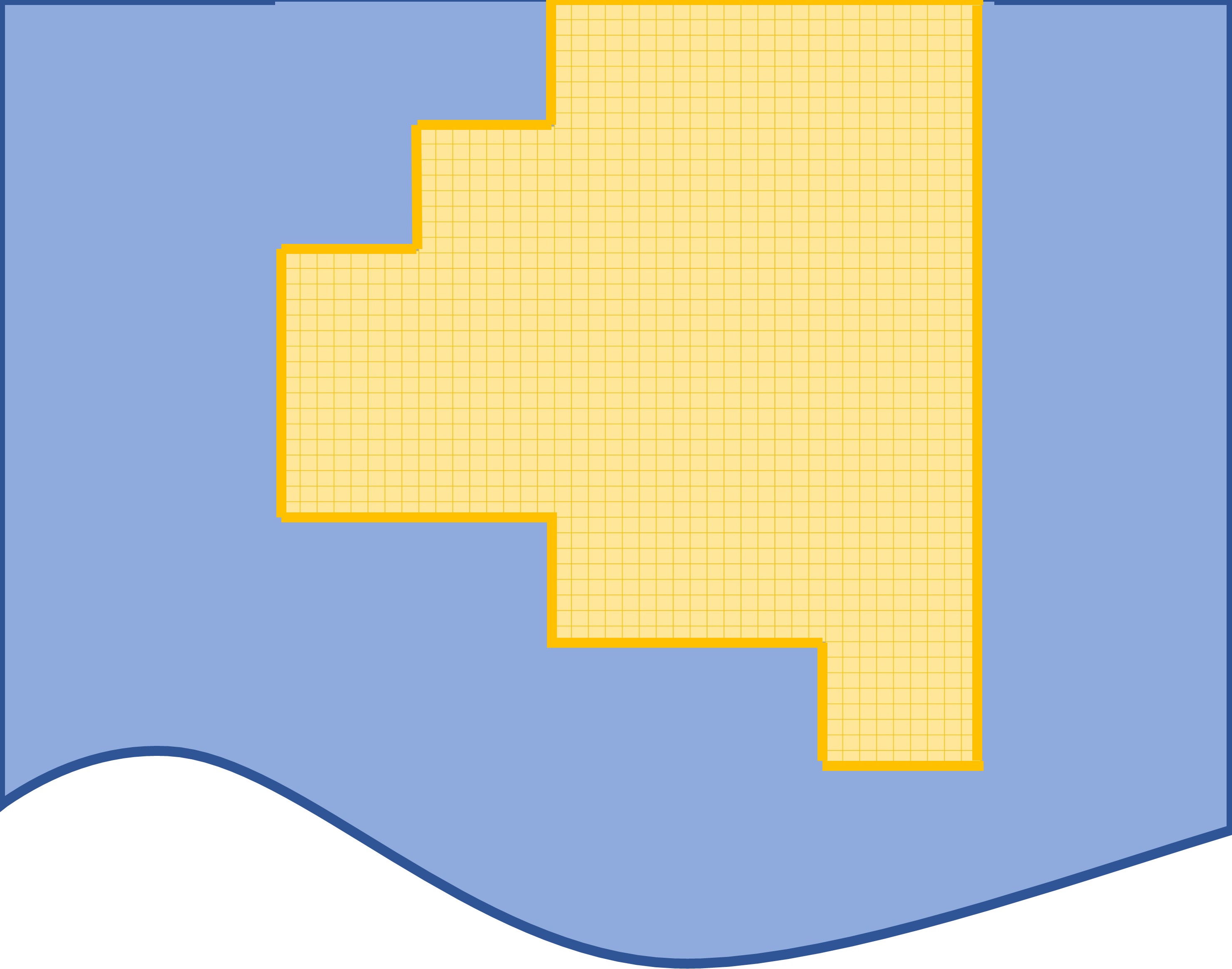}}                                              & -                     &  -     & 0.35 & 0
  \\ \hline\hline
\end{tabular}
\end{table}

\begin{table}[t]
\renewcommand{\arraystretch}{1.5}
\setlength\tabcolsep{5pt}
\small
\centering
\caption{\label{tab:3} Properties of anisotropic inclusion materials. Density $\rho$ (kg/m$^3$), equivalent anisotropy index\cite{sha2018universal,huang2022appraising} $A^{eq}$, elastic constants $c_{ij}$ (GPa), and Voigt averages $c_{11}^0$ and $c_{44}^0$ (GPa).}
\begin{tabular}{llcccccccccccccc}
\hline\hline
& Material & \multirow{1}{*}{$\rho$}  & \multirow{1}{*}{$A^{eq}$} & \multirow{1}{*}{$c_{11}$}  & \multirow{1}{*}{$c_{12}$}  & \multirow{1}{*}{$c_{44}$}  &
\multirow{1}{*}{$c_{11}^0$} &
\multirow{1}{*}{$c_{44}^0$} &\\ \hline
\multirow{3}{*}{Cubic} 
& Aluminum & 2700 & 1.24 & 106.7 & 60.4  & 28.3 & 110.8 & 26.2 \\ 
                       & Inconel  & 8260 & 2.83 & 234.6 & 145.4 & 126.2 & 299.9 & 96.6 \\ 
                       & Lithium  & 534  & 9.14 & 13.4  & 11.3  & 9.6  & 20.4 & 6.18  \\ \hline \hline
                       
& Material & \multirow{1}{*}{$\rho$}  & \multirow{1}{*}{$A^{eq}$} & \multirow{1}{*}{$c_{11}$}  & \multirow{1}{*}{$c_{12}$}  & \multirow{1}{*}{$c_{13}$}  & \multirow{1}{*}{$c_{14}$}  & \multirow{1}{*}{$c_{15}$}  & \multirow{1}{*}{$c_{16}$}  & \multirow{1}{*}{$c_{22}$}  & \multirow{1}{*}{$c_{23}$}  & \multirow{1}{*}{$c_{24}$}  & \multirow{1}{*}{$c_{25}$}  &
\multirow{1}{*}{$c_{26}$}  &\\ \hline
\multirow{3}{*}{Triclinic} & \multirow{3}{*}{CSP} & 2286 & 2.37 & 56.5 & 26.5 & 32.1 & -3.3 & -0.8 & -3.9 & 43.3 & 34.7 & -0.7 & -2.1 & 2.0 \\ \cline{3-15}
&& \multirow{1}{*}{$c_{33}$}  & \multirow{1}{*}{$c_{34}$}  & \multirow{1}{*}{$c_{35}$}  & \multirow{1}{*}{$c_{36}$}  & \multirow{1}{*}{$c_{44}$}  & \multirow{1}{*}{$c_{45}$}  & \multirow{1}{*}{$c_{46}$}  & \multirow{1}{*}{$c_{55}$}  & \multirow{1}{*}{$c_{56}$}  & \multirow{1}{*}{$c_{66}$}  &
\multirow{1}{*}{$c_{11}^0$} &
\multirow{1}{*}{$c_{44}^0$} \\ \cline{3-14}
&& 56.9 & -4.4 & -2.1 & -1.6 & 17.3 & 0.9 & 0.3 & 12.2 & -2.6 & 10.0 & 54.3 & 12.2 \\ \hline\hline
\end{tabular}
\end{table}

The elements on the bottom, left, and right sides of the model are used to define absorbing boundary conditions. The thickness of each absorbing boundary region in the boundary normal direction is chosen to be at least three times the wavelength of the Rayleigh wave in the host material. The absorbing boundary elements in the vicinity of the host material have material properties close to those of the host material but their material damping increases gradually towards the model edge. This gradual increase of damping helps absorb the propagating wave so as to minimise unwanted reflections from the boundaries \cite{rajagopal2012use}. 

The desired Rayleigh wave is generated by applying two sinusoidal time-domain signal of 90$^\circ$ phase shift to multiple source nodes located on the top surface of the model (yellow points in Fig. \ref{fig:1}). The size of the source is set to be equal to three center-frequency wavelengths of the simulated Rayleigh wave, and each source node is assigned a unique amplitude following Eq. (17) in Sarris et al.\cite{sarris2021attenuation}. The simulation is solved using the GPU-accelerated Pogo program \cite{huthwaite2014accelerated} with an explicit time-stepping scheme. A relatively large time step of $\Delta t = 0.9 h /c_L$, satisfying the Courant-Friedrichs-Lewy condition \cite{Gnedin2018}, is used to minimise numerical error \cite{huang2020maximizing}.

Over the course of the FE solution, the $z$ - displacement of the generated incident wave is monitored at a transmitting node (point T in Fig. \ref{fig:1}), while that of the backscattered wave is recorded at a receiving node (point R). We emphasize that the transmitting and receiving nodes are placed respectively far away from the source nodes and the inclusion, in order for the former to monitor the well-formed incident wave and for the latter to record solely the scattered Rayleigh wave in the far field. In addition, a reference signal is obtained at the receiving point using an identical but inclusion-free FE model, and the reference signal is subtracted from the relatively small raw signal to minimise the influence of numerical error. The signal $U_T(t)$ at the transmitting node and the corrected signal $U_R(t)$ at the receiving node are Fourier transformed into the frequency domain to obtain the spectra $U_T(f)$ and $U_R(f)$. The frequency-dependent amplitude of the backscattered Rayleigh wave is then calculated by
\begin{equation}
    U_R(f) = U_T(f) A^\text{sc}(f) \textrm{,}
\end{equation}
which will be used to evaluate the theoretical model result, $A^\text{sc}(\omega)$, in Sec. \ref{sec:4}. It should be noted that this equation is only applicable when the attenuation of the host material and the diffraction losses of the Rayleigh wave are not considered\cite{schmerr2011ultrasonic}.

\begin{figure}[!ht]
\centering
\includegraphics[width=0.9\textwidth]{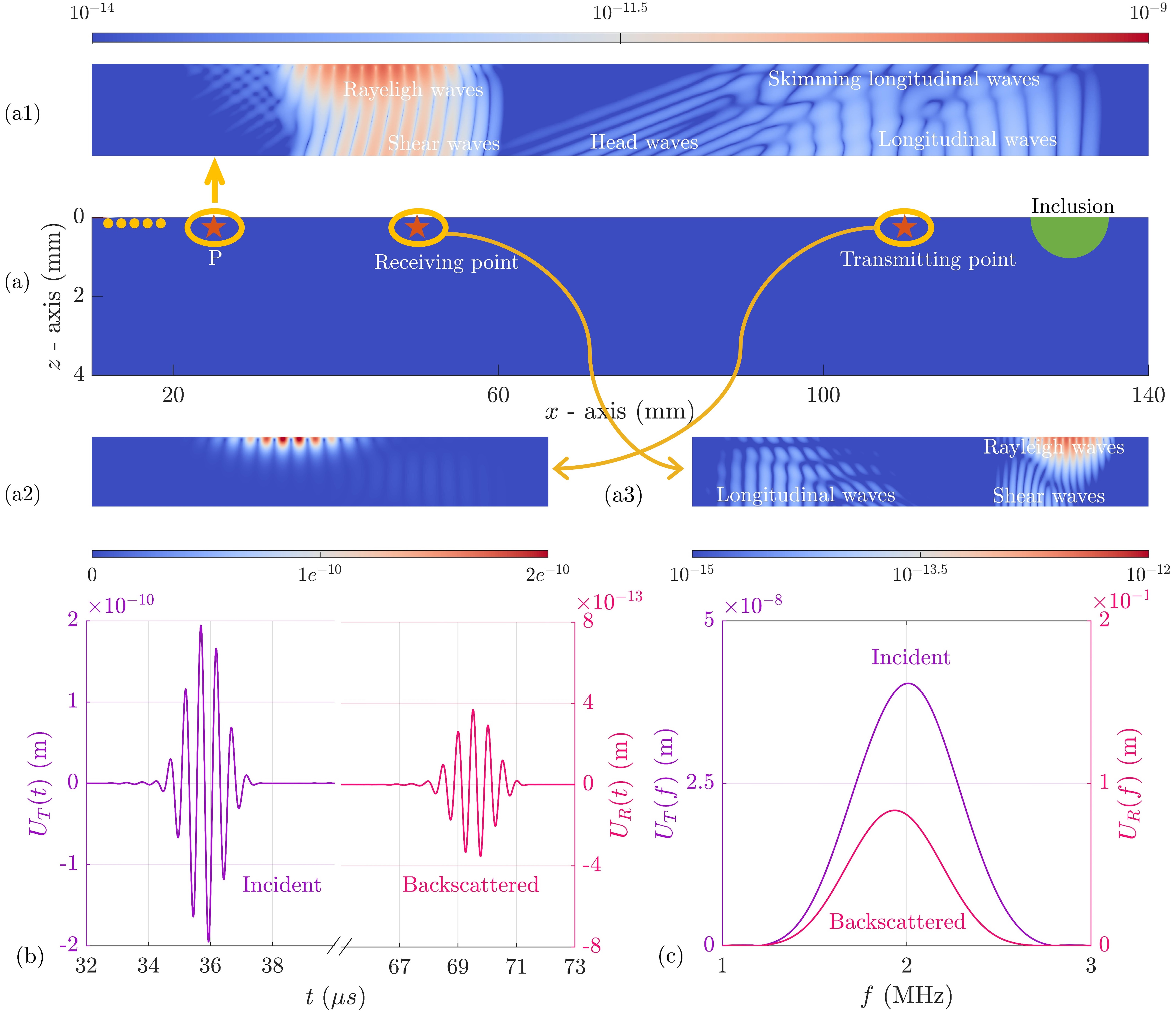}
\caption{\label{fig:2} Example FE modelling of Rayleigh wave scattering by a surface half-circle inclusion. (a) The FE model setup and the simulated wave fields at (a1) the point P at an early time of $t = 10\, \mu s$, (a2) the point close to the inclusion before the incident wave is scattered by the inclusion at $t = 35\,\mu s$ , and (a3) the point far away from the inclusion after the wave scattered by the inclusion at $t = 69\,\mu s$. (b) The $z-$ displacements of the incident Rayleigh wave and the backscattered Rayleigh wave in the time domain. (c)  shows the respective amplitude spectra in the frequency domain.}
\end{figure}

Now we present an example to illustrate the simulated wave field and Rayleigh wave signals. The host material is the aforementioned isotropic aluminum, and the half-circle inclusion (Table \ref{tab:inclusion}) has a 4\% impedance contrast to the host material. The wave field in the model is shown in Fig. \ref{fig:2}(a1) shortly after exciting the source nodes with a signal of 2 MHz center frequency. The wave field shows the coexistence of multiple wave modes in the model, led by faster skimming longitudinal, bulk longitudinal and head waves, and followed by slower bulk shear and Rayleigh waves \cite{hassan2003finite}. However, as the waves propagate further to the transmitting point, the relatively slow Rayleigh wave gets separated from other wave modes and a rather pure Rayleigh wave is formed, as can be seen in Fig. \ref{fig:2}(a2). Similarly, the inclusion causes multiple backscattered wave modes, but as illustrated in Fig. \ref{fig:2}(a3), a relatively pure Rayleigh wave is obtained as the waves reach the receiving point. In this specific case, the scattered Rayleigh wave is about 100 times stronger than the scattered longitudinal wave, confirming our choice of focusing on the scattered Rayleigh wave only in this work. The Rayleigh wave signals recorded at the transmitting and receiving nodes are plotted in Fig. \ref{fig:2}(b), and the respective frequency-domain amplitude spectra are displayed in Fig. \ref{fig:2}(c). The calculated backscattering amplitude will be reported in Sec. \ref{sec:4}.

\section{Results and discussions\label{sec:4}}
\subsection{\label{subsec:4:0}Simple inclusion with different material properties}
It is demonstrated by the theoretical model  in Eq. \ref{eq:016} that both the density difference $\Delta\rho$ and elastic tensor contrast $\Delta c_{pjkl}$ between the inclusion and host materials contribute to Rayleigh wave backscattering. As an essential first step, here we investigate their individual contributions. For this purpose, we consider a simple case of a half-circle inclusion (see Table \ref{tab:inclusion}) on the surface of an aluminum host material. The host material has Young's modulus $E_0$ = $70$ GPa, Poisson's ratio $\nu_0$ = $0.35$ and density $\rho_0$ = 2700 kg/m$^3$. The inclusion is defined in three distinctive cases that differ from the host material in (1) density with $\rho_1$ = 2500 kg/m$^3$, (2) Young's modulus with $E_1$ = $65$ GPa, and (3) both density and Young's modulus with $\rho_1$ = 2600 kg/m$^3$ and $E_1$ = $67$ GPa. These three cases have the same acoustic impedance mismatch of 4\% to the host material. 

\begin{figure}[!ht]
\centering
\includegraphics[width=1\textwidth]{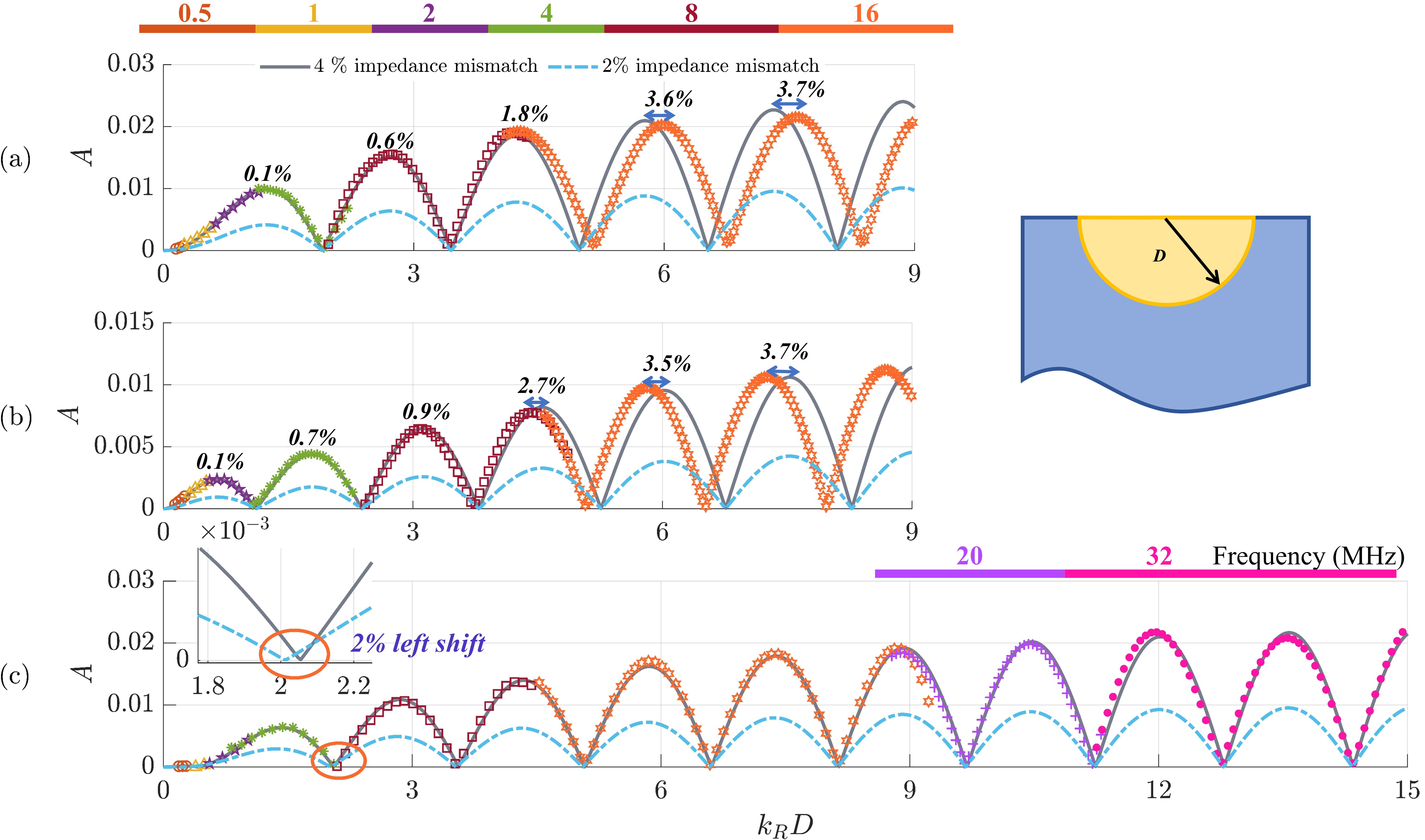}
\caption{\label{fig:3} Rayleigh wave backscattering by a simple half-circle inclusion that differs from the host aluminum in (a) density, (b) Young's modulus and (c) both density and Young's modulus. The resulting backscattering amplitude $A$ is plotted versus normalised frequency $k_R D$ ($D$ is the radius of the inclusion). The theoretical curves calculated from Eq. \ref{eq:016} are compared with the FE points obtained from the models in Table \ref{tab:model} with center frequencies of 0.5, 1, 2, 4, 8 and 16 MHz. The inclusion and host materials have a 4\% impedance mismatch for the solid theoretical lines and FE points, and 2\% mismatch for the dash-dotted theoretical lines. Note that the y-axis range in (b) is half of those in (a) and (c).}
\end{figure}

For these three inclusion cases, the amplitudes of the backscattered Rayleigh waves are plotted in Fig. \ref{fig:3} against the normalized frequency $k_RD$ ($D$ is the radius of the inclusion). The theoretical curves in the figure are calculated from Eq. \ref{eq:016}, while the FE points are simulated using the models in Table \ref{tab:model}. We emphasize that we have achieved a high degree of accuracy for the FE results. A prominent evidence is the overlapping of the points between the neighboring models having different model parameters (thus different numerical errors)\cite{huang2020maximizing}. Therefore, the FE results are well suited to evaluate the approximations of the theoretical model. 

The backscattering results for the three inclusion cases in Fig. \ref{fig:3} demonstrate an oscillating, cyclic behavior. For each case, the peaks of individual cycles increase gradually with $k_RD$, but their cycles are seemingly constant across different cycles. The overall backscattering amplitude for the first case of $\Delta\rho\neq0$ is about twice for the second case of $\Delta c_{pjkl}\neq0$, while the third case resides in the middle. Similarly, the average cycle of the first case is slightly larger than that of the second, again with the third case lying in between. Considering the differences between the three cases, we can infer two main results from the theoretical model in Eq. \ref{eq:016}.

First, $\Delta \rho$ and $\Delta c_{pjkl}$ are scaling factors affecting only the magnitude of backscattering. This is further corroborated by the same cyclic behavior but different amplitudes of the two theoretical curves in each of Fig. \ref{fig:3}(a) and (b); the extra theoretical curve in each plot is obtained using a smaller impedance mismatch of 2\% by varying $\Delta \rho$ or $\Delta c_{pjkl}$.

Second, the remaining integral terms in the equation affect not only the magnitude but also the cyclic period of the backscattering. We can observe from the two cases in Fig. \ref{fig:3}(a) and (b) that, for a given host material and inclusion, the two integral terms associated with $\Delta \rho$ and $\Delta c_{pjkl}$ exhibit different cycles. This is also the reason for the small shift of around 2\% between the two theoretical curves in Fig. \ref{fig:3}(c). In addition, we shall see in the subsection below that the cyclic behavior of the term associated with $\Delta c_{pjkl}$ (presumably for the term associated with $\Delta \rho$ as well) is much more significantly affected by the geometry of the inclusion.

For the first two inclusion cases in Fig. \ref{fig:3}(a) and (b), the theoretical model predictions exhibit very good agreement with the FE results at small $k_R D$. As $k_R D$ increases, the theoretical curve in Fig. \ref{fig:3}(a) tends to have a noticeably shorter cycle (as if compressed) than the FE results, while this trend is reversed for the case in Fig. \ref{fig:3}(b) with the theoretical curve being seemingly expanded. As a result, the agreement between the theoretical and FE results deteriorates as $k_R D$ increases. This is particularly evident as we observe their $k_R D$ differences at individual peaks: the difference increases from 0.1\% at the first peak to 3.7\% at the fifth peak for the first case, and it grows from 0.1\% to 3.5\% at the same peaks for the second case. Interestingly, for the third case with both $\Delta\rho\neq0$ and $\Delta c_{pjkl}\neq0$, the theoretical model agrees well with the FE results even at a very large $k_R D$, with their difference observable only after $k_R D$ $\approx11$.

Such differences can be understood by investigating the phase change caused by each inclusion in comparison to the case where the inclusion is absent. The Born approximation tends to have a larger deviation from the true value when the inclusion-induced phase change gets bigger \cite{kak2001principles, huthwaite2012quantitative}. In our cases, the analytically estimated phase changes are $-0.22 \pi$, $0.20 \pi$ and $0.02 \pi$ for Fig. 3(a), (b) and (c) at $k_R D$ = 9, which are clear evidence supporting the observed large (and similar) theoretical-FE differences in the first two cases and the good agreement in the third one. Apparently, the inclusion-induced phase change increases with frequency, which leads to the increased theoretical-FE difference as observed in each figure panel.

We should emphasize that the three inclusions considered have a smaller density and/or Young's modulus than the host material. We also study the three opposite cases (results not shown) with the density and/or Young's modulus of the inclusion being larger than those of the host. In comparison to Fig. \ref{fig:3}, their FE results are different but their theoretical predictions remain the same (obvious from Eq. \ref{eq:016}); specifically, the theoretical curves are scaled (compressed/expanded over $k_R D$) with respect to the FE results in an opposite way to those in Fig. \ref{fig:3}.

For simplicity, we shall consider only the second case of $\Delta c_{pjkl}\neq0$ in the next two subsections. This is because this particular case involves a considerable approximation in the theoretical solution and is thus beneficial for us to thoroughly evaluate the theoretical model. In this case, the inclusion and host materials will have the same density but different elastic constants.

\subsection{\label{subsec:4:1}Isotropic surface inclusion with different shapes}

\begin{figure}[!ht]
\centering
\includegraphics[width=1\textwidth]{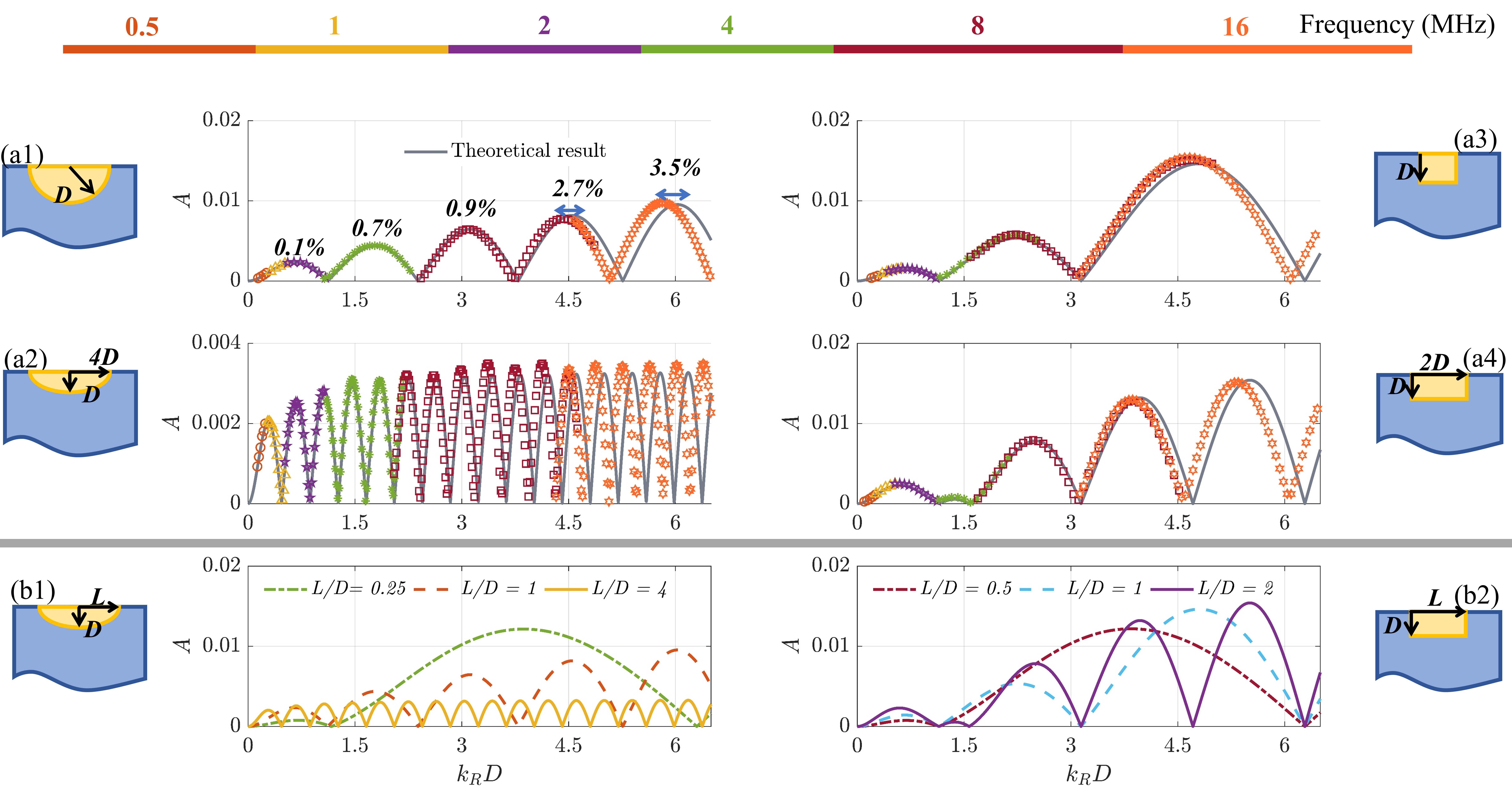}
\caption{\label{fig:4} Rayleigh wave backscattering by a (a1) half-circle , (a2) half-ellipse , (a3) square and (a4) rectangle inclusion. The theoretical results (curves) are compared with the FE results (points) with center frequencies of 0.5, 1, 2, 4, 8, 16 MHz. The inclusion is defined to be slightly different from the host material aluminum only in Young's modulus ($E_1$ = $65$). (b) represents the relationship between the size of inclusion $L/D$ and the cycle of the backscattering amplitude curve for a (b1) half-ellipse and (b2) rectangle inclusion.}
\end{figure}

Now we compare the backscattering amplitudes from the four regularly-shaped inclusions in Table \ref{tab:inclusion}. The host material is aluminum (isotropic and with the above-mentioned material properties) and the inclusion has a 4\% impedance difference to the host caused by $\Delta c_{pjkl}\neq0$. The FE results (points) and theoretical predictions (lines) are displayed in Figs. \ref{fig:4}(a1)-(a4), plotted against $k_R D$. Note that Figure \ref{fig:4}(a1) is the same as Fig. \ref{fig:3}(b). All four cases show a good agreement between the theoretical and numerical results when $k_R D$ is small, which is excellent proof of the validity of the theoretical model.

When comparing Fig. \ref{fig:4}(a1) and (a2) (or (a3) and (a4)), we observe that the results exhibit different cyclic behaviors depending on the lateral dimension of the inclusion. This is further demonstrated in Figures \ref{fig:4}(b1) and (b2), which reveals that the lateral-to-depth dimension ratio $L/D$ affects the cycle of the backscattering amplitude curve and a larger  $L/D$ ratio corresponds to an apparently smaller average cycle for both half-ellipse and rectangle inclusions. This can be explained by that the $z-$ displacement of Rayleigh waves is non-uniform, and the energy of Rayleigh wave is becoming smaller with the increase of depth. Meanwhile, by comparing Fig. \ref{fig:3} with Fig. \ref{fig:4}, it can be seen that the geometry of inclusion has a more obvious effect on the cyclic behaviour.

\begin{figure}[!ht]
\centering
\includegraphics[width=1\textwidth]{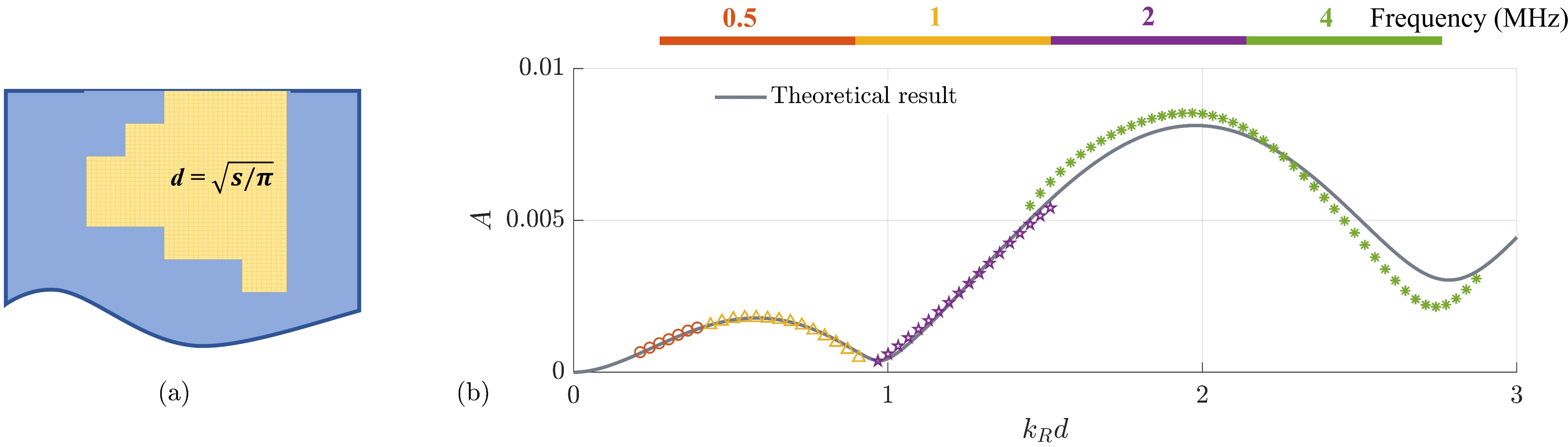}
\caption{\label{fig:5}(a) Irregularly shaped inclusions modeled with uniform square elements; (b) comparison of backscattering amplitude by the irregular inclusion between the FE solution and the theoretical prediction. The host material and inclusion material properties are the same as those of the above regularly shaped inclusion simulations.}
\end{figure}

Then, we investigate the backscattering amplitude from an irregular inclusion (depicted in Fig. \ref{fig:5}(a) and listed in Table \ref{tab:inclusion}). The theoretical prediction and the FE results are compared in Fig. \ref{fig:5}(b), which show very good agreement between each other, demonstrating that the theoretical solution is accurate to describe the backscattered wave of arbitrarily shaped inclusions. As expected from the applicability of the Born approximation, the agreement is becoming worse with the increase of $k_R D$.

\subsection{\label{subsec:4:2} Isotropic subsurface inclusion}

\begin{figure}[!ht]
\centering
\includegraphics[width=1\textwidth]{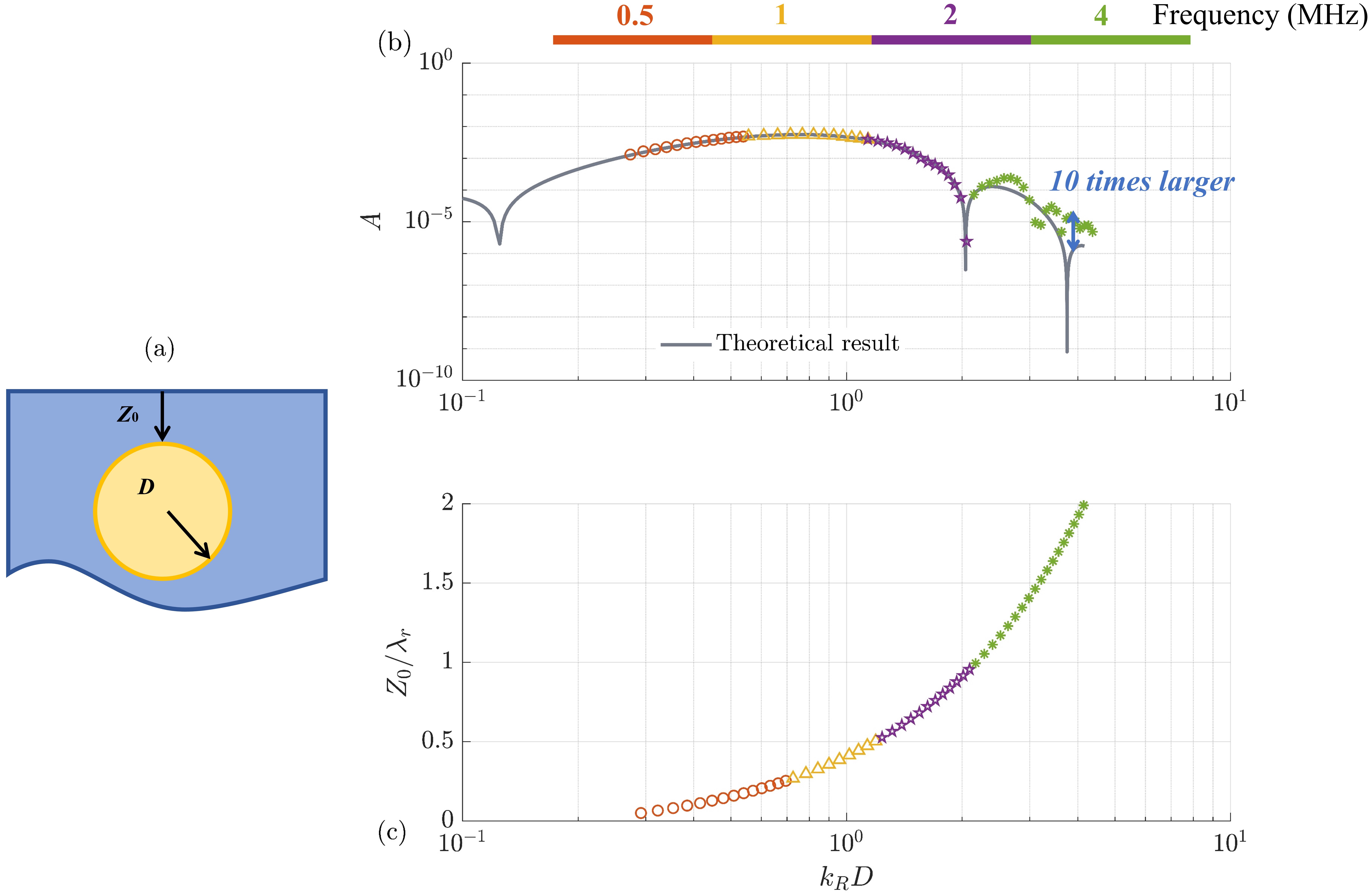}
\caption{\label{fig:6} Rayleigh wave scattering by a subsurface inclusion: (a) illustration of the inclusion; (b) plot of theoretical and FE results versus normalized frequency $k_RD$ on a log scale; (c) the relationship between $Z_0/ \lambda_r$ and frequency. The host material and inclusion material properties are the same as those of the above regularly shaped inclusion simulations.}
\end{figure}

Furthermore, we conduct research to study the backscattering amplitude of a subsurface inclusion. As shown in Fig. \ref{fig:6}(a), we utilize a circular inclusion (parameters in Table \ref{tab:inclusion}) with a depth $Z_0=$ 1.45 mm to the surface of the host material. We set it with $\Delta \rho$ = 0, Young's modulus $E_1$ = 65 GPa and keep other material properties unchanged with respect to the host material.

The analytical result obtained by Eq. \ref{eq:016} is compared with the FE solution in Fig. \ref{fig:6}(b). The results are plotted against the normalized frequency $k_RD$. In the studied frequency range, the ratio of the depth $Z_0$ to the wavelength $\lambda_r$ covers a range of 0 $\sim$ 2 $\lambda_r$, which is displayed in Figure \ref{fig:6}(c). Combining Fig. \ref{fig:6}(b) and (c), we can see that the analytical results match well with the FE results when $Z_0/ \lambda_r$ $<$1, demonstrating a good accuracy of the theoretical model in describing shallow subsurface inclusions. The analytical solution begins to divert from the FE result at around $Z_0$ $\approx$ $\lambda_r$. Their difference increases with $Z_0/\lambda_r$, with the theoretical result being 10 times smaller than the FE results at $Z_0/\lambda_r$  $\approx$ 2, revealing the limitation of the Born approximation for deeper inclusions. Given that the energy of Rayleigh surface waves is generally concentrated in the near-surface region within a depth of about one wavelength \cite{viktrov1967rayleigh}, the limitation can be understandable.

\subsection{\label{subsec:4:3}Anisotropic inclusion}

Here we also evaluate the applicability of our theoretical model to an anisotropic inclusion. This is to identify if the model can be developed further to describe Rayleigh wave scattering in a polycrystalline material in the future. Therefore, for the evaluation, we define the inclusion (half-circle in this case, Table \ref{tab:inclusion}) as a single crystal. Similarly to the polycrystalline material case, we are interested in the average backscattering response from the inclusion when its crystallographic axis is differently oriented, which is exactly what Eq. \ref{eq:020} predicts. In the FE modeling, we randomly rotate the inclusion for 100 times (realizations) and perform FE simulation for each rotated inclusion, and then take the RMS of the 100 backscattering results as the final result. The host material has the same density as the inclusion, and its elastic properties are the Voigt averages of the single inclusion. The single-crystal elastic constants and their Voigt averages are provided in Table \ref{tab:3} for the materials considered.

\begin{figure}[!ht]
\centering
\includegraphics[width=1\textwidth]{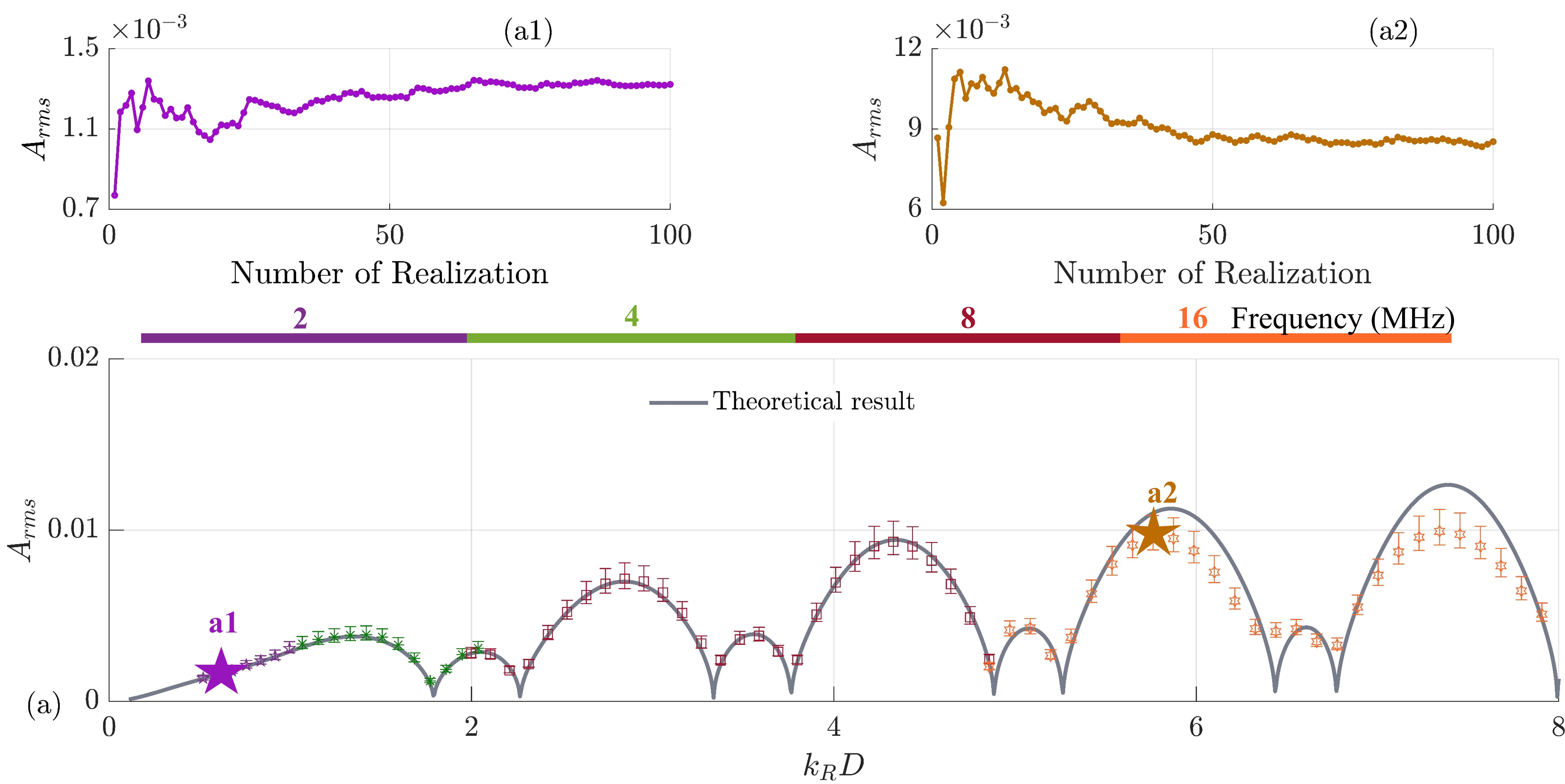}
\caption{\label{fig:7} Rayleigh wave scattering by a half-circle inclusion that is cubic aluminum with an anisotropy index $A^{eq}$ of 1.24. (a) displays the theoretically predicted RMS backscattering amplitude (line) and the FE calculated RMS backscattering amplitude (points) of 100 realizations. The error bar show the 95\% confidence interval\cite{Spiegel2012}. (a1) and (a2) show the convergences of the FE RMS value with the number of realizations at $k_R D$ $\approx$ 0.5 and 5.6.}
\end{figure}

We begin by the case of the inclusion being aluminum, which has a cubic single-crystal symmetry and an anisotropy index of 1.24. The theoretically predicted RMS backscattering amplitudes $A_{rms}$ are plotted as the grey curve in Fig. \ref{fig:7}(a). The respective FE results are plotted as points in the figure, which are the RMS over 100 realizations with the inclusion being randomly rotated in each realization. The error bars show the 95\% confidence interval \cite{Spiegel2012} for the FE points, demonstrating the variation across the realizations with different crystallographic orientations. It is important to note that the use of 100 realizations is sufficient for obtaining statistically converged FE results. This is evidenced by the two convergence curves at $k_R D$ $\approx$ 0.5 and 5.6 (marked as stars in Fig. \ref{fig:7}(a)) in Figs. \ref{fig:7}(a1) and (a2) that show gradually stabilizing RMS value as the number of realizations increases. A general finding from Fig. \ref{fig:7}(a) is that the theoretical and FE results display a good agreement.

\begin{figure}[!ht]
\centering
\includegraphics[width=1\textwidth]{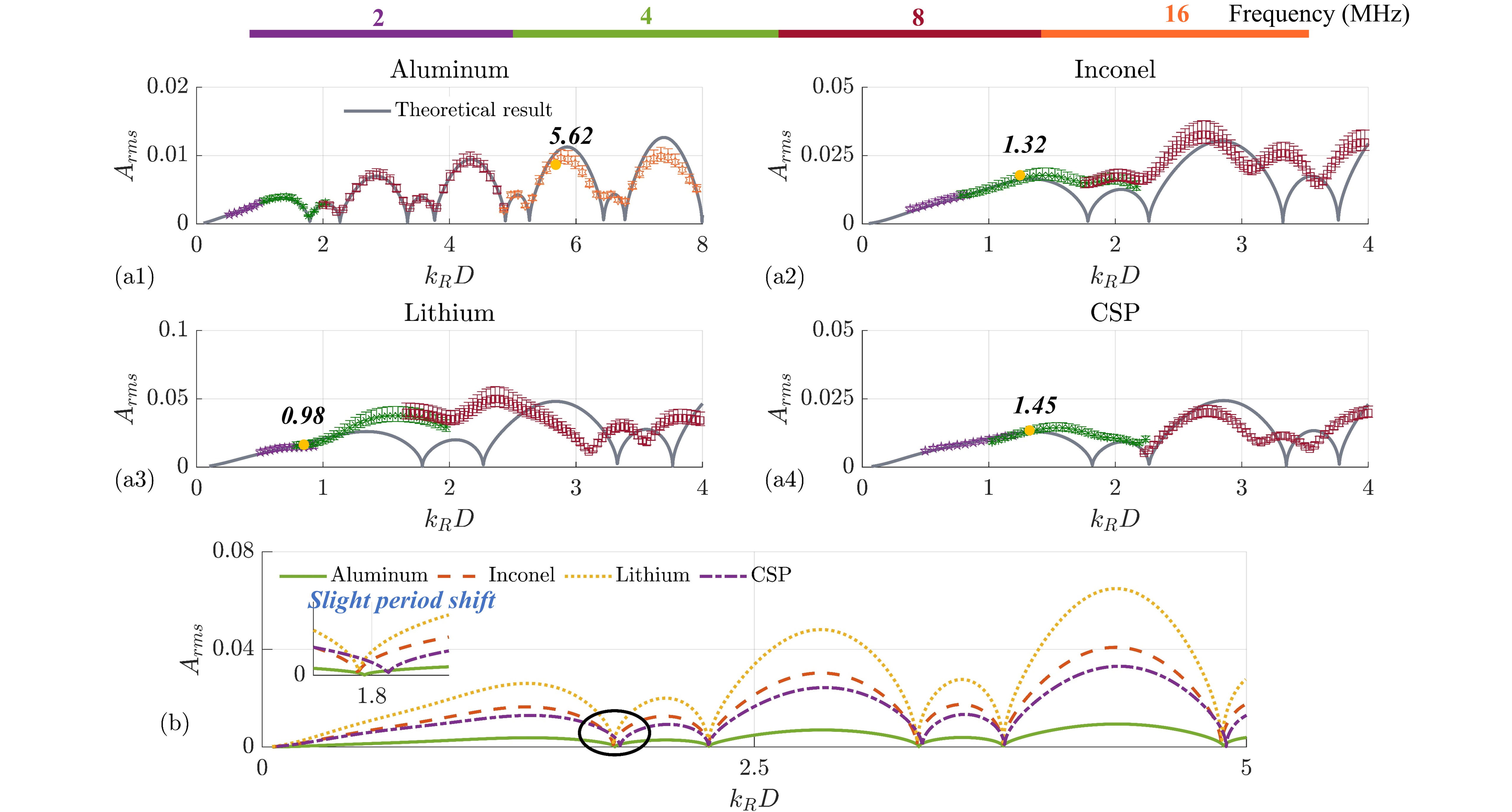}
\caption{\label{fig:8} Rayleigh wave scattering by a simple half-circle inclusion defined as cubic materials of (a1) aluminum ($A^{eq}$ = 1.24), (a2) Inconel ($A^{eq}$ = 2.83) and (a3) lithium ($A^{eq}$ = 9.14), and triclinic material of CSP ($A^{eq}$ = 2.37). The yellow points are the `starting points' where the theoretical and FE results start to deviate. (b) Comparison of the theoretical curves of the four materials.}
\end{figure}

This good agreement is not surprising for an inclusion material (aluminum) of small anisotropy. To observe how the agreement changes with anisotropy, we compare the results of three cubic inclusion materials, namely aluminum, Inconel and lithium, that have anisotropy indices of 1.24, 2.83 and 9.14. Their theoretical and FE results are depicted in Figs. \ref{fig:8}(a1)-(a3). It is clearly demonstrated that the agreement between the theoretical and FE results decreases as anisotropy index increases. This is particularly evident as we observe the `starting point' (exhibited as yellow dots in Fig. \ref{fig:8}) where the theory starts to deviate from the FE results, which is 5.62 for aluminum, 1.32 for Inconel and 0.98 for lithium. Such results are reasonable because the Born approximation is expected to gradually fail with the increase of scattering intensity.

The inclusion materials addressed above are of the highest cubic symmetry. Here we also consider the lowest triclinic symmetry to investigate the influence of single-crystal symmetry. The triclinic material is copper sulfate pentahydrate (CSP) and it has an anisotropy index of 2.37, which is close to that of Inconel. Comparing its results in Fig. \ref{fig:8} (a4) with those of Inconel in Fig. \ref{fig:8} (a3), we can see that the two materials have roughly the same level of theoretical-FE agreement, and their `starting points' are similar. It illustrates that the anisotropy factor has a larger effect on the backscattering amplitude compared with the symmetry of the material.

An extra observation from Figs. \ref{fig:8}(a1)-(a4) is that the cycle of the backscattering amplitude curve only varies subtly among the four materials, as further illustrated in Fig. \ref{fig:8}(b). This supports our earlier finding of the cyclic behavior being mainly determined by the size and shape (rather than the material properties) of the inclusion.

All above results show that the theoretical model can be used to evaluate the backscattering of Rayleigh waves and to establish a direct relation between Rayleigh backscattering and the material and geometrical properties of the inclusion.

\section{\label{sec:5}Conclusion}

In this work, we developed a 2D theoretical model for Rayleigh-to-Rayleigh backscattering by an inclusion. The model is formulated in the frequency domain based on the reciprocity theorem using the far-field Green's function, and the Born approximation is invoked to derive the final result. The model is widely valid for a surface or subsurface inclusion with a regular or irregular geometry, and prominently, the inclusion can be isotropic or anisotropic. A FE model is established to provide relatively accurate reference data for evaluating the approximations of the theoretical model. The comparison of the theoretical and FE results across a range of scattering problems led to various conclusions, mainly including:

1. The theory exhibits very good agreement with the FE results for isotropic inclusions with differently-defined material properties ($\Delta \rho$ or/and $\Delta c_{pjkl}$) and different shapes (half-circle, ellipse, square and rectangle, and even irregular) at small $k_R D$. As a result of the use of the Born approximation, the theory starts to break down gradually with the increase of $k_R D$ due to an increasing phase shift caused by the inclusion. In addition, the sign of the phase shift determines how the theoretical curve scales (compresses or expands over $k_R D$) with respect to the FE results.

2. The theoretical model agrees well with the FE results for the backscattering by a subsurface inclusion when the depth of the inclusion is smaller than the wavelength, and it gradually loses its accuracy as the depth exceeds a wavelength.

3. The theory can predict the RMS backscattering amplitude of Rayleigh waves by an anisotropic inclusion (studied cases include three cubic and one triclinic materials). The results revealed the good applicability of the theory to inclusions of weak anisotropy, and also uncovered the larger effect of the anisotropy, rather than the symmetry, of the inclusion on backscattering. 

4. The backscattering curve demonstrates an oscillating, cyclic behavior. The cycle period is mainly determined by the geometry of the inclusion, and it becomes smaller with the increase of the lateral-to-depth dimension ratio $L/D$ of the inclusion. The material properties of the inclusion and their differences to the host material properties only have a very small influence on the cycle period.

Generally speaking, we have demonstrated the applicability of our theoretical model to a wide spectrum of inclusion types, especially at low frequencies. The model would have great potential for further applications in the area of NDE as well as, for example, for developing models to describe Rayleigh wave scattering in polycrystals.

\section*{Acknowledgments}
This work is supported by the China Scholarship Council and National Natural Science Foundation of China (Grant No. 92060111). BL gratefully acknowledges the Imperial College Research Fellowship and MH the generous funding from the NDE group at Imperial.

\appendix
\section*{Appendices}
\counterwithin{figure}{section}
\counterwithin{table}{section}
\counterwithin{equation}{section}

\section{Detailed backscattering equations}
The expression of $A^\text{sc}\left(\omega \right)$ in Eq. \ref{eq:016} can be further evaluated, leading to
\begin{equation}
    \begin{split}
       A^\text{sc}\left(\omega \right) & =  \sum_{p,j,k,l}^{1,3} A_{pjkl}^\text{sc}\left(\omega \right) = \sum_{l}^{1,3} Aa_{l}^\text{sc} -\sum_{p,j,k,l}^{1,3} Ab_{pjkl}^\text{sc}\left(\omega \right) \textrm{,}
    \end{split}
\end{equation}
with
\begin{linenomath*}
\begin{subequations}
\label{eq:A2}
\allowdisplaybreaks
\begin{align}
 \begin{split}
     Aa^\text{sc}_{1}\left( \omega \right)  =  - A_0  \int_{V}  \Delta \rho \omega^2  U_R^2 \left(\mathbf{x}_{\mathrm{s}}\right)   \exp(2 \text{i} k_R   x_\mathrm{s}) \mathrm{~d} V \textrm{,}
 \end{split}\\
  \begin{split}
     Aa^\text{sc}_{3}\left( \omega \right)  =  ~ A_0  \int_{V}  \Delta \rho \omega^2  W_R^2 \left(\mathbf{x}_{\mathrm{s}}\right)   \exp(2 \text{i} k_R   x_\mathrm{s}) \mathrm{~d} V \textrm{,}
 \end{split}\\
 \begin{split}
     Ab^\text{sc}_{1111}\left( \omega \right)  =  ~ A_0 \Delta c_{1111}\left(\mathbf{x}_{\mathrm{s}} \right) \int_{V} \left [k_R^2   U_{R}(z_\mathrm{s})^2 \right  ] \exp \left ( 2 \text{i} k_R   x_\mathrm{s} \right)  \mathrm{~d} V \textrm{,}
 \end{split}\\
 \begin{split}
 Ab^\text{sc}_{1113}\left( \omega \right)  =  ~ A_0 \Delta c_{1113}\left(\mathbf{x}_{\mathrm{s}} \right) \int_{V}  \left [ \text{i} k_R^2   U_{R}(z_\mathrm{s}) W_{R}(z_\mathrm{s}))   \right ]  \exp \left ( 2 \text{i} k_R   x_\mathrm{s} \right)  \mathrm{~d} V \textrm{,}
 \end{split}\\
 \begin{split}
     Ab^\text{sc}_{1131}\left(  \omega \right)  =  ~  A_0 \Delta c_{1131}\left(\mathbf{x}_{\mathrm{s}} \right) \int_{V} \left [-\text{i} k_R  U_{R}(z_\mathrm{s}) \frac{\partial U_{R} \left(z_\mathrm{s}\right)}{\partial z_\mathrm{s}} \right   ]  \exp \left ( 2 \text{i} k_R   x_\mathrm{s} \right)  \mathrm{~d} V \textrm{,}
 \end{split}\\
 \begin{split}
   Ab^\text{sc}_{1133}\left( \omega \right)  =  ~A_0 \Delta c_{1133}\left(\mathbf{x}_{\mathrm{s}} \right) \int_{V} \left [k_R U_{R}(z_\mathrm{s}) \frac{\partial W_{R} \left(z_\mathrm{s}\right)} {\partial z_\mathrm{s}}   \right ]  \exp \left ( 2 \text{i} k_R  x_\mathrm{s} \right)  \mathrm{~d} V \textrm{,}
 \end{split}\\
 \begin{split}
    Ab^\text{sc}_{1313}\left(\omega \right)  =  ~ A_0 \Delta c_{1313}\left(\mathbf{x}_{\mathrm{s}} \right) \int_{V} \left [   - k_R^2 W_{R}(z_\mathrm{s})   W_{R}(z_\mathrm{s})      \right ]  \exp \left ( 2 \text{i} k_R   x_\mathrm{s} \right)  \mathrm{~d} V \textrm{,}
 \end{split}\\
 \begin{split}
     Ab^\text{sc}_{1331}\left( \omega \right)  =  ~ A_0 \Delta c_{1331}\left(\mathbf{x}_{\mathrm{s}} \right) \int_{V} \left [k_R W_{R}(z_\mathrm{s}) \frac{\partial U_{R} \left(z_\mathrm{s}\right)} {\partial z_\mathrm{s}}       \right ]  \exp \left ( 2 \text{i} k_R  x_\mathrm{s} \right)  \mathrm{~d} V \textrm{,}
 \end{split}\\
 \begin{split}
     Ab^\text{sc}_{1333}\left( \omega \right)  =  ~ A_0 \Delta c_{1333}\left(\mathbf{x}_{\mathrm{s}} \right) \int_{V} \left [ \text{i} k_R  W_{R}(z_\mathrm{s}) \frac{\partial W_{R} \left(z_\mathrm{s}\right)}{\partial z_\mathrm{s}}        \right ]  \exp \left ( 2 \text{i} k_R  x_\mathrm{s} \right)  \mathrm{~d} V \textrm{,}
 \end{split}\\
 \begin{split}
     Ab^\text{sc}_{3131}\left(  \omega \right) =  ~  A_0 \Delta c_{1331}\left(\mathbf{x}_{\mathrm{s}} \right) \int_{V} \left [-\frac{\partial U_{R} \left( z_\mathrm{s} \right) }{\partial z_\mathrm{s}}  \frac{\partial U_{R} \left(z_\mathrm{s}\right)} {\partial z_\mathrm{s}}      \right ]   \exp \left ( 2 \text{i} k_R   x_\mathrm{s} \right)  \mathrm{~d} V \textrm{,} 
 \end{split}\\
 \begin{split}
     Ab^\text{sc}_{3133}\left(  \omega \right) =  ~  A_0 \Delta c_{1333}\left(\mathbf{x}_{\mathrm{s}} \right) \int_{V} \left [ -\text{i} \frac{\partial U_{R} \left( z_\mathrm{s} \right) }{\partial z_\mathrm{s}}  \frac{\partial W_{R} \left(z_\mathrm{s}\right)} {\partial z_\mathrm{s}}     \right ]   \exp \left ( 2 \text{i} k_R   x_\mathrm{s} \right)  \mathrm{~d} V \textrm{,}
 \end{split}\\
 \begin{split}
     Ab^\text{sc}_{3333}\left( \omega \right) =  ~  A_0 \Delta c_{3333}\left(\mathbf{x}_{\mathrm{s}} \right) \int_{V} \left [ \frac{\partial W_{R} \left(z_\mathrm{s}\right)}{\partial z_\mathrm{s}} \frac{\partial W_{R} \left(z_\mathrm{s}\right)}{\partial z_\mathrm{s}}    \right ]   \exp \left (2 \text{i} k_R   x_\mathrm{s} \right)  \mathrm{~d} V \textrm{,} 
\end{split}
\\
\begin{split}
Ab^\text{sc}_{1311}\left( \omega \right)
 = Ab^\text{sc}_{1113}\left( \omega \right),
 Ab^\text{sc}_{3111}\left( \omega \right)
 = Ab^\text{sc}_{1131}\left( \omega \right), 
Ab^\text{sc}_{3113}\left( \omega \right)
 = Ab^\text{sc}_{1331}\left( \omega \right),   \\
Ab^\text{sc}_{3311}\left( \omega \right)
 = Ab^\text{sc}_{1133}\left( \omega \right), 
Ab^\text{sc}_{3313}\left( \omega \right)
 = Ab^\text{sc}_{1333}\left( \omega \right), 
Ab^\text{sc}_{3331}\left( \omega \right)
 = Ab^\text{sc}_{3133}\left( \omega \right), 
 \end{split}
\end{align}
\end{subequations}
\end{linenomath*}

The detailed expression of the RMS of the backscattering amplitude $A^\text{sc}_{rms}\left( \omega \right)$ in Eq. \ref{eq:020} can be written as
\begin{equation}\label{eq:A3}
    \begin{split}
       A^\text{sc}_{rms}\left(\omega \right) & = \sqrt{\left< A_{pjkl}^\text{sc}\left(\omega \right) A^{sc}_{\alpha\beta\gamma\delta}\left(\omega \right)\right>} = \sqrt{\left< \sum _{pjkl}^{1, 3} Ab_{pjkl}^\text{sc}\left(\omega \right) \sum_{\alpha\beta\gamma\delta}^{1, 3} Ab^{sc}_{\alpha\beta\gamma\delta}\left(\omega \right) \right>} \\
    & = \sqrt{\left<Ab_{1111}^\text{sc}Ab_{1111}^\text{sc}\right>+\left<Ab_{1111}^\text{sc}Ab_{1113}^\text{sc}\right>+\left<Ab_{1111}^\text{sc}Ab_{1131}^\text{sc}\right>+...+\left<Ab_{3333}^\text{sc}Ab_{3333}^\text{sc}\right>} \textrm{,}
    \end{split}
\end{equation}
where the individual terms in the square root are obtained by multiplying and averaging the respective $A$ terms in Eq. \ref{eq:A2}. Here, we give an example expression for one of the terms by
\begin{equation}
\allowdisplaybreaks
\begin{split}
\left<Ab_{1111}^\text{sc}Ab_{1111}^\text{sc}\right>  =  A_0^2 & \left<\Delta c_{1111}\Delta c_{1111}\right>  \\ & \times \int_{V} \int_{V} k_R^4  U_{R}(z)^2 U_{R}(z_\mathrm{s})^2  \exp \left ( 2 \text{i} k_R   (x+x_\mathrm{s}) \right)  \mathrm{~d} V \mathrm{~d} V \textrm{.}
\end{split}
\end{equation}

The elastic covariance $\left<\Delta c_{pjkl}\Delta c_{\alpha\beta\gamma\delta}\right>$ involved in Eq. \ref{eq:A3} is given by
\begin{equation}
    \begin{split}
       & \left<\Delta c_{pjkl}\Delta c_{\alpha\beta\gamma\delta}\right>  = \left< c_{pjkl} c_{\alpha\beta\gamma\delta}\right>- \left< c_{pjkl} \right> \left< c_{\alpha\beta\gamma\delta}\right> \\
        & = \left(\left< a_{ja} a_{pb} a_{kc} a_{ld} a_{\alpha m} a_{\beta n} a_{\gamma o} a_{\delta q} \right>- \left< a_{ja} a_{pb} a_{kc} a_{ld} \right> \left< a_{\alpha m} a_{\beta n} a_{\gamma o} a_{\delta q} \right> \right) c_{abcd}c_{mnoq}    \end{split} 
\end{equation}
for materials of any symmetry. $\textbf{a} (\theta, \phi, \xi)$ is the rotation matrix defined using the Euler angles \cite{weaver1990diffusivity}. The final expressions are provided in the supplementary material.

\clearpage
\bibliography{Reference}

\end{document}

\setstretch{1.5}
% BEGIN SUPPLEMENTARY MATERIAL 
\clearpage
\pagebreak

\begin{center}
    \textbf{{\large Supplemental Material for: ``Theoretical and numerical modeling of Rayleigh wave scattering by an elastic inclusion''}}\\[1em]
    Shan Li, Ming Huang, Yongfeng Song, Bo Lan, Xiongbing Li
\end{center}

For any symmetry material:

$\left<\Delta c_{11}\Delta c_{15}\right>$ = $\left<\Delta c_{11}\Delta c_{35}\right>$ = $\left<\Delta c_{13}\Delta c_{15}\right>$ = $\left<\Delta c_{15}\Delta c_{55} \right>$ = 0;

$\left<\Delta c_{13}\Delta c_{35}\right>$ = $\left<\Delta c_{33}\Delta c_{15} \right>$ = $\left<\Delta c_{33}\Delta c_{35}\right>$ = $\left<\Delta c_{35}\Delta c_{55} \right>$ = 0;

$\left<\Delta c_{11}\Delta c_{11}\right>$ = $\left<\Delta c_{33}\Delta c_{33}\right>$; 

$\left<\Delta c_{11}\Delta c_{13}\right>$ = $\left<\Delta c_{13}\Delta c_{33}\right>$;
\begin{equation}
        \begin{split}
           \left<\Delta c_{11} \Delta c_{11}\right >  = & (112c_{11}^2 + 16c_{11}c_{12} + 32c_{12}^2 + 16c_{11}c_{13} - 
     16c_{12}c_{13} + 32c_{13}^2 + 80c_{14}^2 \\&
     + 400c_{15}^2 + 400c_{16}^2 - 
     96c_{11}c_{22} + 16c_{12}c_{22} - 
          64c_{13}c_{22} + 112c_{22}^2 
          - 64c_{11}c_{23} \\& 
          - 16c_{12}c_{23} - 
     16c_{13}c_{23} + 16c_{22}c_{23} + 32c_{23}^2 + 160c_{14}c_{24} + 400c_{24}^2 + 160c_{15}c_{25} \\&
     + 80c_{25}^2 + 
          480c_{16}c_{26} + 400c_{26}^2 - 96c_{11}c_{33} - 64c_{12}c_{33} + 
     16c_{13}c_{33} - 96c_{22}c_{33} \\&+ 16c_{23}c_{33} + 112c_{33}^2 + 160c_{14}c_{34} + 
     480c_{24}c_{34} + 400c_{34}^2 + 
          480c_{15}c_{35} + 160c_{25}c_{35} \\&+ 400c_{35}^2 + 160c_{16}c_{36} + 
     160c_{26}c_{36} + 80c_{36}^2 - 128c_{11}c_{44} - 32c_{12}c_{44} - 32c_{13}c_{44} \\&+ 
     32c_{22}c_{44} + 
          128c_{23}c_{44} + 32c_{33}c_{44} + 128c_{44}^2 + 320c_{16}c_{45} + 
     320c_{26}c_{45} + 320c_{36}c_{45} \\&+ 320c_{45}^2 + 320c_{15}c_{46} + 
     320c_{25}c_{46} + 320c_{35}c_{46} + 
          320c_{46}^2 + 32c_{11}c_{55} - 32c_{12}c_{55} \\&+ 128c_{13}c_{55} - 
     128c_{22}c_{55} - 32c_{23}c_{55} + 32c_{33}c_{55} - 64c_{44}c_{55} + 128c_{55}^2 + 
     320c_{14}c_{56} \\&+ 
          320c_{24}c_{56} + 320c_{34}c_{56} + 320c_{56}^2 + 32c_{11}c_{66} + 
     128c_{12}c_{66} - 32c_{13}c_{66} + 32c_{22}c_{66} \\&- 32c_{23}c_{66} - 
     128c_{33}c_{66} - 64c_{44}c_{66} - 
          64c_{55}c_{66} + 128c_{66}^2)/1575;  
        \end{split}
    \end{equation}
 \\
\begin{equation}
    \begin{split}
        \left<\Delta c_{11} \Delta c_{13} \right> & = (4c_{11}^2 + 32c_{11}c_{12} + 4c_{12}^2 + 32c_{11}c_{13} - 12c_{12}c_{13} + 
     4c_{13}^2 + 20c_{14}^2 
     \\&- 20c_{15}^2 - 20c_{16}^2 - 12c_{11}c_{22} + 
     32c_{12}c_{22} -48c_{13}c_{22} + 4c_{22}^2 - 48c_{11}c_{23} 
     \\&
     - 12c_{12}c_{23} - 
     12c_{13}c_{23} + 32c_{22}c_{23} + 4c_{23}^2 + 160c_{14}c_{24} - 20c_{24}^2 + 
     160c_{15}c_{25} 
     \\& 
     + 20c_{25}^2 + 
          120c_{16}c_{26} - 20c_{26}^2 - 12c_{11}c_{33} - 48c_{12}c_{33} + 
     32c_{13}c_{33} - 12c_{22}c_{33} 
     \\&
     + 32c_{23}c_{33} + 4c_{33}^2 + 160c_{14}c_{34} + 
     120c_{24}c_{34} - 20c_{34}^2 + 
          120c_{15}c_{35} + 160c_{25}c_{35} 
          \\&- 20c_{35}^2 + 160c_{16}c_{36} + 
     160c_{26}c_{36} + 20c_{36}^2 + 24c_{11}c_{44} - 4c_{12}c_{44} - 4c_{13}c_{44} \\&+ 
     4c_{22}c_{44} - 24c_{23}c_{44} + 
          4c_{33}c_{44} - 64c_{44}^2 - 40c_{16}c_{45} - 40c_{26}c_{45} - 
     40c_{36}c_{45} 
     \\&- 160c_{45}^2 - 40c_{15}c_{46} - 40c_{25}c_{46} - 40c_{35}c_{46} - 
     160c_{46}^2 + 4c_{11}c_{55} - 
          4c_{12}c_{55} 
          \\&- 24c_{13}c_{55} + 24c_{22}c_{55} - 4c_{23}c_{55} + 
     4c_{33}c_{55} + 32c_{44}c_{55} - 64c_{55}^2 - 40c_{14}c_{56} 
     \\&- 40c_{24}c_{56} - 
     40c_{34}c_{56} - 160c_{56}^2 + 
          4c_{11}c_{66} - 24c_{12}c_{66} - 4c_{13}c_{66} + 4c_{22}c_{66} 
    \\& - 
     4c_{23}c_{66} + 24c_{33}c_{66} + 32c_{44}c_{66} + 32c_{55}c_{66} - 64c_{66}^2)/
   1575; 
    \end{split}
\end{equation}

\begin{equation}
    \begin{split}
       \left<\Delta c_{11} \Delta c_{33} \right> = & (-48c_{11}^2 - 24c_{11}c_{12} + 2c_{12}^2 - 24c_{11}c_{13} + 4c_{12}c_{13} + 
     2c_{13}^2 - 120c_{15}^2 
     \\& - 120c_{16}^2 + 54c_{11}c_{22} - 24c_{12}c_{22} + 
     36c_{13}c_{22} - 
          48c_{22}^2 + 36c_{11}c_{23} + 4c_{12}c_{23} \\& + 4c_{13}c_{23} - 
     24c_{22}c_{23} + 2c_{23}^2 - 120c_{14}c_{24} - 120c_{24}^2 - 120c_{15}c_{25} - 
     360c_{16}c_{26} \\&
     - 120c_{26}^2 + 
          54c_{11}c_{33} + 36c_{12}c_{33} - 24c_{13}c_{33} + 54c_{22}c_{33} - 
     24c_{23}c_{33} - 48c_{33}^2 \\&- 120c_{14}c_{34} - 360c_{24}c_{34} - 120c_{34}^2 - 
     360c_{15}c_{35} - 120c_{25}c_{35} - 
          120c_{35}^2 \\&- 120c_{16}c_{36} 
          - 120c_{26}c_{36} + 72c_{11}c_{44} + 
     8c_{12}c_{44} + 8c_{13}c_{44} - 48c_{22}c_{44} + 8c_{23}c_{44}\\& - 48c_{33}c_{44} + 
     8c_{44}^2 - 240c_{16}c_{45} - 
          240c_{26}c_{45} - 240c_{15}c_{46} - 240c_{35}c_{46} - 48c_{11}c_{55} \\&+ 
     8c_{12}c_{55}  + 8c_{13}c_{55} + 72c_{22}c_{55} + 8c_{23}c_{55} - 48c_{33}c_{55} +      16c_{44}c_{55} + 8c_{55}^2 - 
          240c_{24}c_{56} \\&- 240c_{34}c_{56} - 48c_{11}c_{66} + 8c_{12}c_{66} + 
     8c_{13}c_{66} - 48c_{22}c_{66} + 8c_{23}c_{66} + 72c_{33}c_{66} \\&+ 16c_{44}c_{66} + 
     16c_{55}c_{66} + 8c_{66}^2)/1575;  
    \end{split}
\end{equation}
\begin{equation}
    \begin{split}
       \left<\Delta c_{11} \Delta c_{55} \right> = & (4c_{11}^2 + 2c_{11}c_{12} - 16c_{12}^2 + 2c_{11}c_{13} + 8c_{12}c_{13} - 
     16c_{13}^2 - 40c_{14}^2 \\&- 20c_{15}^2 - 20c_{16}^2 - 12c_{11}c_{22} + 
     2c_{12}c_{22} + 12c_{13}c_{22} + 
          4c_{22}^2 + 12c_{11}c_{23} \\&+ 8c_{12}c_{23} + 8c_{13}c_{23} + 2c_{22}c_{23} - 
     16c_{23}^2 - 20c_{14}c_{24} - 20c_{24}^2 - 20c_{15}c_{25} \\&- 40c_{25}^2 + 
     120c_{16}c_{26} - 20c_{26}^2 - 
          12c_{11}c_{33} + 12c_{12}c_{33} + 2c_{13}c_{33} - 12c_{22}c_{33} \\&+ 
     2c_{23}c_{33} + 4c_{33}^2 - 20c_{14}c_{34} + 120c_{24}c_{34} - 20c_{34}^2 + 
     120c_{15}c_{35} - 20c_{25}c_{35} \\&- 
          20c_{35}^2 - 20c_{16}c_{36} - 20c_{26}c_{36} - 40c_{36}^2 - 
     36c_{11}c_{44} + 6c_{12}c_{44} + 6c_{13}c_{44} \\&+ 34c_{22}c_{44} - 44c_{23}c_{44} + 
     34c_{33}c_{44} - 24c_{44}^2 + 
          140c_{16}c_{45} + 140c_{26}c_{45} - 100c_{36}c_{45} \\&- 40c_{45}^2 + 
     140c_{15}c_{46} - 100c_{25}c_{46} + 140c_{35}c_{46} - 40c_{46}^2 + 
     34c_{11}c_{55} + 6c_{12}c_{55} \\&- 44c_{13}c_{55} - 
          36c_{22}c_{55} + 6c_{23}c_{55} + 34c_{33}c_{55} - 8c_{44}c_{55} - 
     24c_{55}^2 - 100c_{14}c_{56} \\&+ 140c_{24}c_{56} + 140c_{34}c_{56} - 40c_{56}^2 + 
     34c_{11}c_{66} - 44c_{12}c_{66} + 
          6c_{13}c_{66} + 34c_{22}c_{66} \\&+ 6c_{23}c_{66} - 36c_{33}c_{66} - 
     8c_{44}c_{66} - 8c_{55}c_{66} - 24c_{66}^2)/1575;     \end{split}
\end{equation}

\begin{equation}
    \begin{split}
     \left<\Delta c_{13} \Delta c_{13} \right> = & (8c_{11}^2 + 4c_{11}c_{12} + 83c_{12}^2 + 4c_{11}c_{13} - 74c_{12}c_{13} + 
     83c_{13}^2 + 240c_{14}^2 \\&+ 60c_{15}^2 + 60c_{16}^2 + c_{11}c_{22} + 
     4c_{12}c_{22} - 26c_{13}c_{22} + 
          8c_{22}^2 - 26c_{11}c_{23} \\&- 74c_{12}c_{23} - 74c_{13}c_{23} + 
     4c_{22}c_{23} + 83c_{23}^2 + 60c_{14}c_{24} + 60c_{24}^2 + 60c_{15}c_{25} \\&+ 
     240c_{25}^2 - 60c_{16}c_{26} + 
          60c_{26}^2 + c_{11}c_{33} - 26c_{12}c_{33} + 4c_{13}c_{33} + c_{22}c_{33} \\&+ 
     4c_{23}c_{33} + 8c_{33}^2 + 60c_{14}c_{34} - 60c_{24}c_{34} + 60c_{34}^2 - 
     60c_{15}c_{35} + 60c_{25}c_{35} \\&+ 
          60c_{35}^2 + 60c_{16}c_{36} + 60c_{26}c_{36} + 240c_{36}^2 + 
     28c_{11}c_{44} + 52c_{12}c_{44} + 52c_{13}c_{44} \\&- 32c_{22}c_{44} - 68c_{23}c_{44} - 
     32c_{33}c_{44} + 92c_{44}^2 - 
          120c_{16}c_{45} - 120c_{26}c_{45} - 240c_{36}c_{45} \\&+ 240c_{45}^2 - 
     120c_{15}c_{46} - 240c_{25}c_{46} - 120c_{35}c_{46} + 240c_{46}^2 - 
     32c_{11}c_{55} + 52c_{12}c_{55} \\&- 
          68c_{13}c_{55} + 28c_{22}c_{55} + 52c_{23}c_{55} - 32c_{33}c_{55} - 
     56c_{44}c_{55} + 92c_{55}^2 - 240c_{14}c_{56} \\&- 120c_{24}c_{56} - 
     120c_{34}c_{56} + 240c_{56}^2 - 32c_{11}c_{66} - 
          68c_{12}c_{66} + 52c_{13}c_{66} - 32c_{22}c_{66} \\&+ 52c_{23}c_{66} + 
     28c_{33}c_{66} - 56c_{44}c_{66} - 56c_{55}c_{66} + 92c_{66}^2)/1575; 
       \end{split}
\end{equation}

\begin{equation}
    \begin{split}
        \left<\Delta c_{13} \Delta c_{55} \right> = & (8c_{11}^2 - 6c_{11}c_{12} - 17c_{12}^2 - 6c_{11}c_{13} + 26c_{12}c_{13} - 
     17c_{13}^2 - 60c_{14}^2 \\&+ 60c_{15}^2 + 60c_{16}^2 + c_{11}c_{22} - 
     6c_{12}c_{22} - 6c_{13}c_{22} + 
          8c_{22}^2 - 6c_{11}c_{23}\\& + 26c_{12}c_{23} + 26c_{13}c_{23} - 6c_{22}c_{23} - 
     17c_{23}^2 + 60c_{24}^2 - 60c_{25}^2 - 60c_{16}c_{26} \\&+ 60c_{26}^2 + 
     c_{11}c_{33} - 6c_{12}c_{33} - 
          6c_{13}c_{33} + c_{22}c_{33} - 6c_{23}c_{33} + 8c_{33}^2 \\&- 60c_{24}c_{34} + 
     60c_{34}^2 - 60c_{15}c_{35} + 60c_{35}^2 - 60c_{36}^2 + 8c_{11}c_{44} - 
     38c_{12}c_{44} \\&- 38c_{13}c_{44} - 
          22c_{22}c_{44} + 112c_{23}c_{44} - 22c_{33}c_{44} + 12c_{44}^2 - 
     60c_{16}c_{45} \\&- 60c_{26}c_{45} + 300c_{36}c_{45} - 60c_{15}c_{46} + 
     300c_{25}c_{46} - 60c_{35}c_{46} - 22c_{11}c_{55} \\& -38c_{12}c_{55} + 112c_{13}c_{55} + 8c_{22}c_{55} - 38c_{23}c_{55} - 
     22c_{33}c_{55} + 24c_{44}c_{55} \\& + 12c_{55}^2 + 300c_{14}c_{56} - 60c_{24}c_{56} - 
     60c_{34}c_{56} - 22c_{11}c_{66} + 
          112c_{12}c_{66} \\& - 38c_{13}c_{66} - 22c_{22}c_{66} - 38c_{23}c_{66} + 
     8c_{33}c_{66} + 24c_{44}c_{66} + 24c_{55}c_{66} + 12c_{66}^2)/1575; 
    \end{split}
\end{equation}

\begin{equation}
    \begin{split}
      \left<\Delta c_{15} \Delta c_{15} \right> = &  (25c_{11}^2 - 5c_{11}c_{12} + 15c_{12}^2 - 5c_{11}c_{13} - 5c_{12}c_{13} + 
     15c_{13}^2 + 35c_{14}^2 \\&+ 115c_{15}^2 + 115c_{16}^2 - 15c_{11}c_{22} - 
     5c_{12}c_{22} - 
          10c_{13}c_{22} + 25c_{22}^2 - 10c_{11}c_{23} \\&- 5c_{12}c_{23} - 
     5c_{13}c_{23} - 5c_{22}c_{23} + 15c_{23}^2 + 10c_{14}c_{24} + 115c_{24}^2 + 
     10c_{15}c_{25} \\&+ 35c_{25}^2 + 
          30c_{16}c_{26} + 115c_{26}^2 - 15c_{11}c_{33} - 10c_{12}c_{33} - 
     5c_{13}c_{33} - 15c_{22}c_{33} \\&- 5c_{23}c_{33} + 25c_{33}^2 + 10c_{14}c_{34} + 
     30c_{24}c_{34} + 115c_{34}^2 + 
          30c_{15}c_{35} + 10c_{25}c_{35} \\&+ 115c_{35}^2 + 10c_{16}c_{36} + 
     10c_{26}c_{36} + 35c_{36}^2 - 20c_{11}c_{44} - 10c_{12}c_{44} - 10c_{13}c_{44} \\&- 
     10c_{22}c_{44} + 60c_{23}c_{44} - 
          10c_{33}c_{44} + 60c_{44}^2 + 20c_{16}c_{45} + 20c_{26}c_{45} + 
     140c_{36}c_{45} \\&+ 140c_{45}^2 + 20c_{15}c_{46} + 140c_{25}c_{46} + 
     20c_{35}c_{46} + 140c_{46}^2 - 10c_{11}c_{55} - 
          10c_{12}c_{55} \\&+ 60c_{13}c_{55} - 20c_{22}c_{55} - 10c_{23}c_{55} - 
     10c_{33}c_{55} - 20c_{44}c_{55} + 60c_{55}^2 + 140c_{14}c_{56} \\&+ 20c_{24}c_{56} + 
     20c_{34}c_{56} + 140c_{56}^2 - 
          10c_{11}c_{66} + 60c_{12}c_{66} - 10c_{13}c_{66} - 10c_{22}c_{66} \\&- 
     10c_{23}c_{66} - 20c_{33}c_{66} - 20c_{44}c_{66} - 20c_{55}c_{66} + 60c_{66}^2)/
   1575; 
     \end{split}
\end{equation}
\\
\begin{equation}
    \begin{split}
       \left<\Delta c_{15} \Delta c_{35} \right> = & (30c_{11}^2 + 30c_{11}c_{12} - 15c_{12}^2 + 30c_{11}c_{13} - 15c_{13}^2 - 
     30c_{14}^2 \\&+ 30c_{15}^2 + 30c_{16}^2 - 45c_{11}c_{22} + 30c_{12}c_{22} - 
     30c_{13}c_{22} + 
          30c_{22}^2 \\&- 30c_{11}c_{23} + 30c_{22}c_{23} - 15c_{23}^2 + 
     120c_{14}c_{24} + 30c_{24}^2 + 120c_{15}c_{25} \\&- 30c_{25}^2 + 360c_{16}c_{26} + 
     30c_{26}^2 - 45c_{11}c_{33} - 
          30c_{12}c_{33} + 30c_{13}c_{33} \\&- 45c_{22}c_{33} + 30c_{23}c_{33} + 
     30c_{33}^2 + 120c_{14}c_{34} + 360c_{24}c_{34} + 30c_{34}^2 \\&+ 360c_{15}c_{35} + 
     120c_{25}c_{35} + 30c_{35}^2 + 
          120c_{16}c_{36} + 120c_{26}c_{36} \\&- 30c_{36}^2 - 60c_{11}c_{44} + 
     60c_{22}c_{44} - 60c_{23}c_{44} + 60c_{33}c_{44} - 60c_{44}^2 \\&+ 240c_{16}c_{45} + 
     240c_{26}c_{45} - 120c_{36}c_{45} - 
          120c_{45}^2 + 240c_{15}c_{46} - 120c_{25}c_{46} \\&+ 240c_{35}c_{46} - 
     120c_{46}^2 + 60c_{11}c_{55} - 60c_{13}c_{55} - 60c_{22}c_{55} + 60c_{33}c_{55} \\&- 
     60c_{55}^2 - 120c_{14}c_{56} + 
          240c_{24}c_{56} + 240c_{34}c_{56} - 120c_{56}^2 + 60c_{11}c_{66} \\&- 
     60c_{12}c_{66} + 60c_{22}c_{66} - 60c_{33}c_{66} - 60c_{66}^2)/3150;  
    \end{split}
\end{equation}

\begin{equation}
    \begin{split}
     \left<\Delta  c_{35} \Delta c_{35} \right> = & (50c_{11}^2 - 10c_{11}c_{12} + 30c_{12}^2 - 10c_{11}c_{13} - 
     10c_{12}c_{13} + 30c_{13}^2 \\&+ 70c_{14}^2 + 230c_{15}^2 + 230c_{16}^2 - 
     30c_{11}c_{22} - 10c_{12}c_{22} - 
          20c_{13}c_{22} \\&+ 50c_{22}^2 - 20c_{11}c_{23} - 10c_{12}c_{23} - 
     10c_{13}c_{23} - 10c_{22}c_{23} + 30c_{23}^2 \\&+ 20c_{14}c_{24} + 230c_{24}^2 + 
     20c_{15}c_{25} + 70c_{25}^2 + 
          60c_{16}c_{26} + 230c_{26}^2 \\&- 30c_{11}c_{33} - 20c_{12}c_{33} - 
     10c_{13}c_{33} - 30c_{22}c_{33} - 10c_{23}c_{33} + 50c_{33}^2 \\&+ 20c_{14}c_{34} + 
     60c_{24}c_{34} + 230c_{34}^2 + 
          60c_{15}c_{35} + 20c_{25}c_{35} + 230c_{35}^2 \\&+ 20c_{16}c_{36} + 
     20c_{26}c_{36} + 70c_{36}^2 - 40c_{11}c_{44} - 20c_{12}c_{44} - 20c_{13}c_{44} \\&- 
     20c_{22}c_{44} + 120c_{23}c_{44} - 
          20c_{33}c_{44} + 120c_{44}^2 + 40c_{16}c_{45} + 40c_{26}c_{45} \\&+ 
     280c_{36}c_{45} + 280c_{45}^2 + 40c_{15}c_{46} + 280c_{25}c_{46} + 
     40c_{35}c_{46} + 280c_{46}^2 \\&- 20c_{11}c_{55} - 
          20c_{12}c_{55} + 120c_{13}c_{55} - 40c_{22}c_{55} - 20c_{23}c_{55} - 
     20c_{33}c_{55}\\& - 40c_{44}c_{55} + 120c_{55}^2 + 280c_{14}c_{56} + 40c_{24}c_{56} + 
     40c_{34}c_{56} + 280c_{56}^2 \\&- 
          20c_{11}c_{66} + 120c_{12}c_{66} - 20c_{13}c_{66} - 20c_{22}c_{66} - 
     20c_{23}c_{66} - 40c_{33}c_{66} \\& - 40c_{44}c_{66} - 40c_{55}c_{66} + 
     120c_{66}^2)/3150;    
    \end{split}
\end{equation}

\begin{equation}
    \begin{split}
\left<\Delta c_{33} \Delta c_{55} \right> = & (4c_{11}^2 + 2c_{11}c_{12} - 16c_{12}^2 + 2c_{11}c_{13} + 8c_{12}c_{13} - 
     16c_{13}^2 \\&- 40c_{14}^2 - 20c_{15}^2 - 20c_{16}^2 - 12c_{11}c_{22} + 
     2c_{12}c_{22} + 12c_{13}c_{22} \\&+ 
          4c_{22}^2 + 12c_{11}c_{23} + 8c_{12}c_{23} + 8c_{13}c_{23} + 2c_{22}c_{23} - 
     16c_{23}^2 \\&- 20c_{14}c_{24} - 20c_{24}^2 - 20c_{15}c_{25} - 40c_{25}^2 + 
     120c_{16}c_{26} - 20c_{26}^2 \\&- 
          12c_{11}c_{33} + 12c_{12}c_{33} + 2c_{13}c_{33} - 12c_{22}c_{33} + 
     2c_{23}c_{33} + 4c_{33}^2 \\&- 20c_{14}c_{34} + 120c_{24}c_{34} - 20c_{34}^2 + 
     120c_{15}c_{35} - 20c_{25}c_{35} - 
          20c_{35}^2 \\&- 20c_{16}c_{36} - 20c_{26}c_{36} - 40c_{36}^2 - 
     36c_{11}c_{44} + 6c_{12}c_{44} + 6c_{13}c_{44} \\&+ 34c_{22}c_{44} - 44c_{23}c_{44} + 
     34c_{33}c_{44} - 24c_{44}^2 + 
          140c_{16}c_{45} + 140c_{26}c_{45} \\&- 100c_{36}c_{45} - 40c_{45}^2 + 
     140c_{15}c_{46} - 100c_{25}c_{46} + 140c_{35}c_{46} - 40c_{46}^2 \\&+ 
     34c_{11}c_{55} + 6c_{12}c_{55} - 44c_{13}c_{55} -           36c_{22}c_{55} + 6c_{23}c_{55} + 34c_{33}c_{55} \\&- 8c_{44}c_{55} - 
     24c_{55}^2 - 100c_{14}c_{56} + 140c_{24}c_{56} + 140c_{34}c_{56} - 40c_{56}^2 +\\& 
     34c_{11}c_{66} - 44c_{12}c_{66} + 
          6c_{13}c_{66} + 34c_{22}c_{66} + 6c_{23}c_{66} - 36c_{33}c_{66} \\&- 
     8c_{44}c_{66} - 8c_{55}c_{66} - 24c_{66}^2)/1575;      \end{split}
\end{equation}

\begin{equation}
    \begin{split}
       \left<\Delta c_{55} \Delta c_{55} \right> = &  (8c_{11}^2 - 16c_{11}c_{12} + 23c_{12}^2 - 16c_{11}c_{13} - 14c_{12}c_{13} + 
     23c_{13}^2 \\&+ 60c_{14}^2 + 60c_{15}^2 + 60c_{16}^2 + c_{11}c_{22} - 
     16c_{12}c_{22} + 
          14c_{13}c_{22} \\&+ 8c_{22}^2 + 14c_{11}c_{23} - 14c_{12}c_{23} - 
     14c_{13}c_{23} - 16c_{22}c_{23} + 23c_{23}^2 \\&- 60c_{14}c_{24} + 60c_{24}^2 - 
     60c_{15}c_{25} + 60c_{25}^2 - 
          60c_{16}c_{26} + 60c_{26}^2 \\&+ c_{11}c_{33} + 14c_{12}c_{33} - 16c_{13}c_{33} + 
     c_{22}c_{33} - 16c_{23}c_{33} + 8c_{33}^2 \\&- 60c_{14}c_{34} - 60c_{24}c_{34} + 
     60c_{34}^2 - 60c_{15}c_{35} - 
          60c_{25}c_{35} + 60c_{35}^2 \\&- 60c_{16}c_{36} - 60c_{26}c_{36} + 
     60c_{36}^2 - 12c_{11}c_{44} + 12c_{12}c_{44} + 12c_{13}c_{44} \\&- 12c_{22}c_{44} + 
     12c_{23}c_{44} - 12c_{33}c_{44} + 
          72c_{44}^2 + 180c_{45}^2 + 180c_{46}^2 \\&- 12c_{11}c_{55} +      12c_{12}c_{55} + 12c_{13}c_{55} - 12c_{22}c_{55} + 12c_{23}c_{55} - 12c_{33}c_{55} \\&- 
     36c_{44}c_{55} + 72c_{55}^2 + 
          180c_{56}^2 - 12c_{11}c_{66} + 12c_{12}c_{66} + 12c_{13}c_{66} \\&- 
     12c_{22}c_{66} + 12c_{23}c_{66} - 12c_{33}c_{66} - 36c_{44}c_{66} - 36c_{55}c_{66} + 
     72c_{66}^2)/1575;  
    \end{split}
\end{equation}